\documentclass{llncs}

\usepackage[T1]{fontenc}

\usepackage{multirow}

\usepackage{graphicx}
\usepackage{subfigure}
\usepackage{epstopdf}
\usepackage{algorithm,algpseudocode}
\usepackage{tikz}
\usepackage{pgfplots}
\usepackage{amsmath}

\usepackage{comment}

\usepackage{lscape}

\usepackage{hyperref}

\begin{document}
\sloppy
\title{TPGen: A Self-Stabilizing GPU-Based Method for Prime and Test Paths Generation}
%\title{Contribution Title\thanks{Supported by organization x.}}
\author{Ebrahim Fazli\inst{1} \and  Ali Ebnenasir\inst{2}}
%
%\authorrunning{E.  Fazli and A. Ebnenasir}
%\titlerunning{TPGen: A Self-Stabilizing GPU-Based Method for Prime and Test Paths Generation}

\institute{Department of Computer Engineering,Islamic Azad University, Zanjan, Iran \\
\email{efazli@znu.ac.ir}\\
 \and
Michigan Technological University, Houghton MI 49931, USA\\
\email{aebnenas@mtu.edu}}

%\institute{Princeton University, Princeton NJ 08544, USA \and
%Springer Heidelberg, Tiergartenstr. 17, 69121 Heidelberg, Germany
\maketitle              % typeset the header of the contribution
\begin{abstract}
This paper presents a novel scalable GPU-based method for Test Paths (TPs) and Prime Paths (PPs)  Generation, called TPGen,   used in structural testing and in test data generation.  TPGen outperforms existing methods for PPs and TPs generation in several orders of magnitude, both in time and space efficiency.  Improving  both time and space efficiency is made possible through devising a new non-contiguous and hierarchical memory allocation method, called Three-level Path Access Method (TPAM), that enables efficient storage of maximal simple paths in memory.  In addition to its high time and space efficiency,  a major significance of TPGen includes its self-stabilizing design where threads execute in a fully asynchronous and order-oblivious way  without using any atomic instructions.  TPGen can generate PPs and TPs of structurally complex programs that have an extremely high cyclomatic and Npath complexity. 

% Another contribution of this paper includes a novel notion of structural complexity defined based on PPs that can distinguish complex program structures that remian undetected by existing compexity criteria. %

%\keywords{Structural Testing \and Prime Path  \and Test Path \and GPU Programming \and Self-stabilization.}
\end{abstract}

\section{Introduction}
\label{sec:Intro}

This paper presents a scalable GPU-based method for the Generation of all Test  Paths (TPs) and Prime Paths (PPs), called TPGen,  for structural testing.   Complete Path Coverage (CPC) is an ideal testing requirement where all execution paths in a program are tested.  However,  such coverage may be impossible because some execution paths may be infeasible,  and the total number of program paths may be unbounded due to loops and recursion.  Lowering expectations,  one would resort to testing all simple paths, where no vertex is repeated in a simple path,  but the Control Flow Graph (CFG) of even small programs may have an extremely large number of simple paths.  Amman and Offut  \cite{ammann2016introduction}  propose the notion of Prime Path Coverage (PPC), where a {\it prime path} is a {\em maximal} simple path; a simple path that is not included in any other simple path.  PP coverage is an important testing requirement as it subsumes other coverage criteria (e.g., branch coverage) in structural testing.  As such,  finding the set of all PPs of a program (1) expands the scope of path coverage, and (2) enables the generation of Test Paths (TPs),  which are very important in test data generation.  This paper presents a scalable approach for the generation of PPs and TPs in structurally complex programs. %, as well as introducing a path complexity criterion based on PPs.%

%Problem: extracting prime and test path in structurally complex programs based on Npath and PP complexity measure.

%As the NPath complexity grows exponentially, the time cost grows linearly.

%Path coverage in structural testing: Ideally, we would like to test all execution paths,  called Complete Path Coverage (CPC). However, this problem is in general undecidable. \cite{??}.  Lowering expectations, one would resort to testing all simple paths, where no vertex is repeated in a simple path.  However,  the Control Flow Graph (CFG) of even small programs may have extremely large number of simple paths. Amman and Offut  \cite{ammann2016introduction}  propose the notion of Prime Path Coverage (PPC), where a a{\it prime path} is a maximal simple path; a simple path that is not subsumed by any other simple path. 

Despite the crucial role of PPC in structural testing, there are a limited number of methods that offer effective and efficient algorithms for generating PPs and TPs for complex real-world programs.  Amman and Offutt \cite{ammann2008graph}  propose  a dynamic programming solution for  extracting all PPs.  
Dwarakanath and Jankiti \cite{dwarakanath2014minimum} utilize Max-Flow/Min-Cut algorithms to generate minimum number of TPs that cover all PPs.  Hoseini and Jalili \cite{hoseini2014automatic}  use genetic algorithms to generate PPs/TPs of CFGs  extracted from sequential programs.   Sayyari and Emadi \cite{sayyari2015automated} exploit  ant colony algorithms to generate TPs covering PPs.  Sirvastava {\it et al.} \cite{srivastava2010optimized} extract a Markov chain model and produce an optimal test set.  Bidgoli {\it et al.} \cite{bidgoli2017using} apply swarm intelligence algorithms using a normalized fitness function to ensure the coverage of PPs.  Lin and Yeh \cite{lin2001automatic} and also Bueno and Jino \cite{bueno2002automatic}  present methods based on genetic algorithm to cover PPs. Our previous work  \cite{fazli2019time} generates PPs and TPs  in a  compositional fashion where we separately extract the PPs of each Strongly Connected Component (SCC) of a Program's Control Flow Graph (CFG),   and then merge them towards generating the PPs of the CFG.  
Most aforementioned methods are applicable to toy examples and small programs and cannot be utilized for PP coverage of  programs that have a high structural complexity; i.e.,  very large number of PPs.  This paper exploits the processing power of GPUs in order to provide a time and space efficient parallel algorithm for the generation of all PPs. %We also introduce a novel notion of structural complexity, called {\it Prime Path Complexity} which clearly distinguishes distinct program structures and provides a realistic measure of complexity for path coverage.  %

%Most of the existing methods in the field of PPs/TPs generation are practically unsuccessful in dealing with large real-world programs that are structurally complex. There are several reasons for this failure.  First, the number of PPs of these programs is very large. Second, The entire processing of the large CFGs with a large number of simple paths is extremely time and memory consuming, due to the data structure, contiguous memory allocation and memory management methods associated with this process. Third, owing to not using the information obtained in the PPs extracting phase in the TPs generation phase, most of the existing methods have a high cost of reprocessing time.

{\bf Contributions}: The major contributions of this paper are multi-fold. First,  we present a novel high-performance GPU-based algorithm for PPs and TPs generation that works in a self-stabilizing fashion.  The TPGen algorithm first generates the component graph of the input CFG on the CPU and then processes each vertex of that component graph (each SCC) in parallel on a GPU.  TPGen is  vertex-based in that each thread $T_i$ is mapped to a vertex $v_i$ and a list $l_i$ of partial paths is associated with $v_i$.  Each thread extends  the paths in $l_i$ while ensuring their simplicity.   The execution of threads is completely asynchronous.  In each iteration,  thread $T_i$ updates $l_i$ based on the extension of the paths in the predecessors of $v_i$,  and removes all covered simple paths from $l_i$.  The experimental evaluations of TPGen show that it can generate all PPs of programs with extremely large cyclomatic \cite{mccabe1976complexity} and Npath complexity \cite{nejmeh1988npath} in a time and space efficient way.  Specifically,  TPGen outperforms existing sequential methods up to 3.5 orders of magnitude in terms of  time efficiency  and up to 2 orders of magnitude in space efficiency for a given benchmark.  TPGen achieves such efficiency while ensuring data race-freedom without using `atomic' statements in its design.  Moreover,  TPGen is self-stabilizing in the sense that the threads start in any order and generate PPs without any kind of synchronization with each other. Such lack of synchronization significantly improves time efficiency but is hard to design due to risk of thread interferences.  As a result,  we consider the  design of TPGen as a model for other GPU-based algorithms, which by itself is a novel contribution.   
Second, we propose a non-contiguous and hierarchical memory allocation method, called Three-level Path Access Method (TPAM), that enables efficient storage of maximal simple paths in memory. Last but not least, we put forward a benchmark of synthetic programs for evaluating the structural complexity of programs and for experimental evaluation of PPs/TPs generation methods.

\textbf {Organization}. 
Section \ref{sec:prelim} defines some basic concepts. Section \ref{sec:ProblemStatement} states the PPs generation problem. Subsequently, 
Section \ref{sec:dataStruct} presents the TPAM method of memory allocation.  
Section \ref{sec:parAlg} puts forward a highly time and space-efficient parallel algorithm implemented on GPU for PPs generation.
Section \ref{sec:ExperimantalResults} presents our experimental results. Section \ref{sec:RelWorks} discusses related work. 
Finally, Section \ref{sec:conclude} makes concluding remarks and discusses future extensions of this work.

%Section \ref{sec:prelim} defines some basic concepts related to the directed graphs, PPs and TPs. Section \ref{sec:ProblemStatement} states the PPs generation problem. Subsequently, Section \ref{sec:ParallelMainMethod} puts forward a highly time and space-efficient method based on CPU-GPU for parallel generation of PPs/TPs paths. Section \ref{sec:ExperimantalResults} talks about our experimental results. Section \ref{sec:RelWorks} introduces background of PPC and classifies related works. Finally, Section \ref{sec:conclude} makes concluding remarks and discusses future extensions of this work.

\section{Preliminaries}
\label{sec:prelim}

\begin{figure}[!ht]
%\Figure[h]{
  \centering
\subfigure[java.util.Arrays.binarySearch0()]{

  \includegraphics[scale = .35]{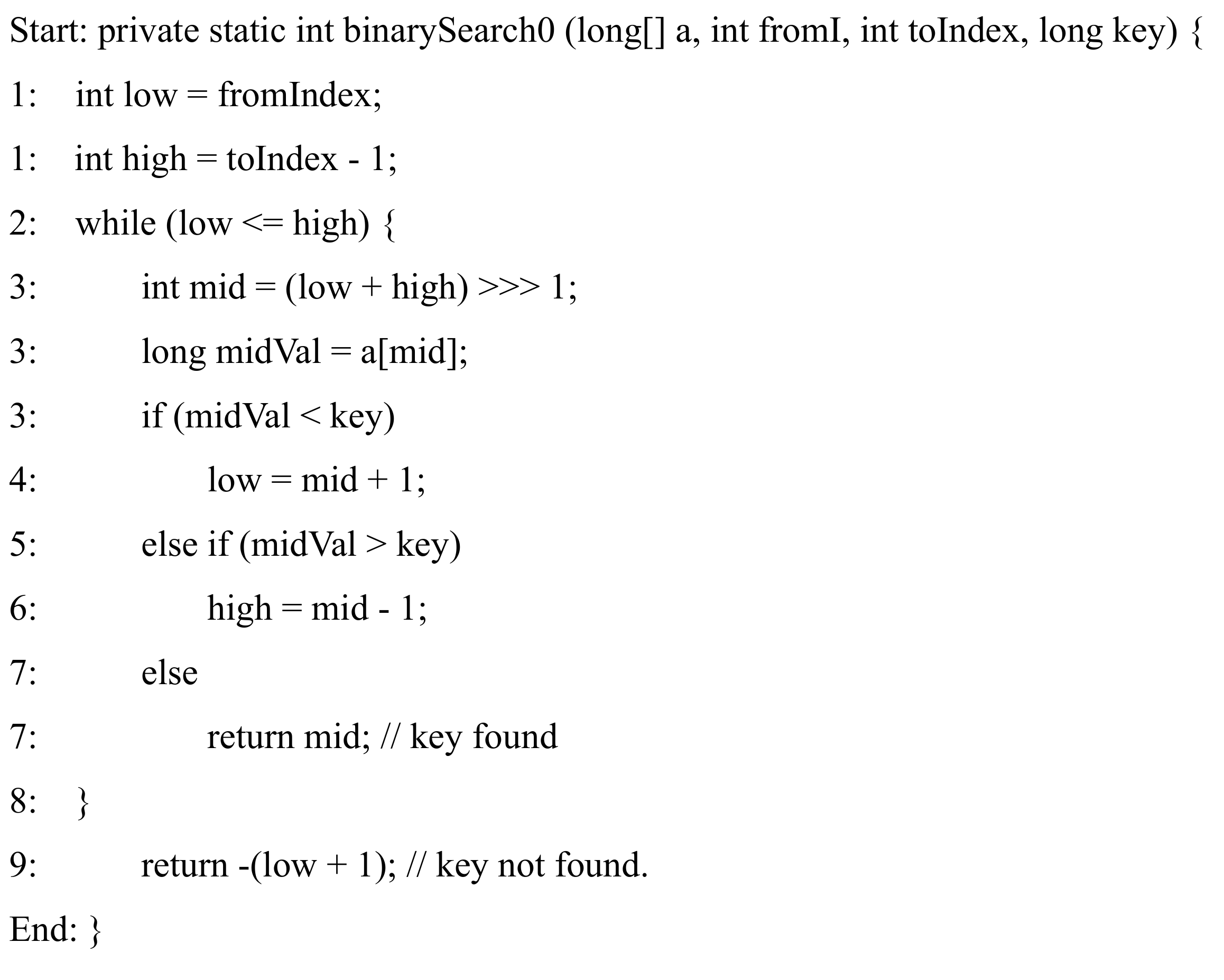}
%  \caption{java.util.Arrays.rangeCheck()}
  \label{fig:ExampleMethod}
%\end{subfigure}%
}
\hspace*{0.0cm}
\subfigure[CFG for method (a) ]{
  %\centering
  \includegraphics[scale= 0.4]{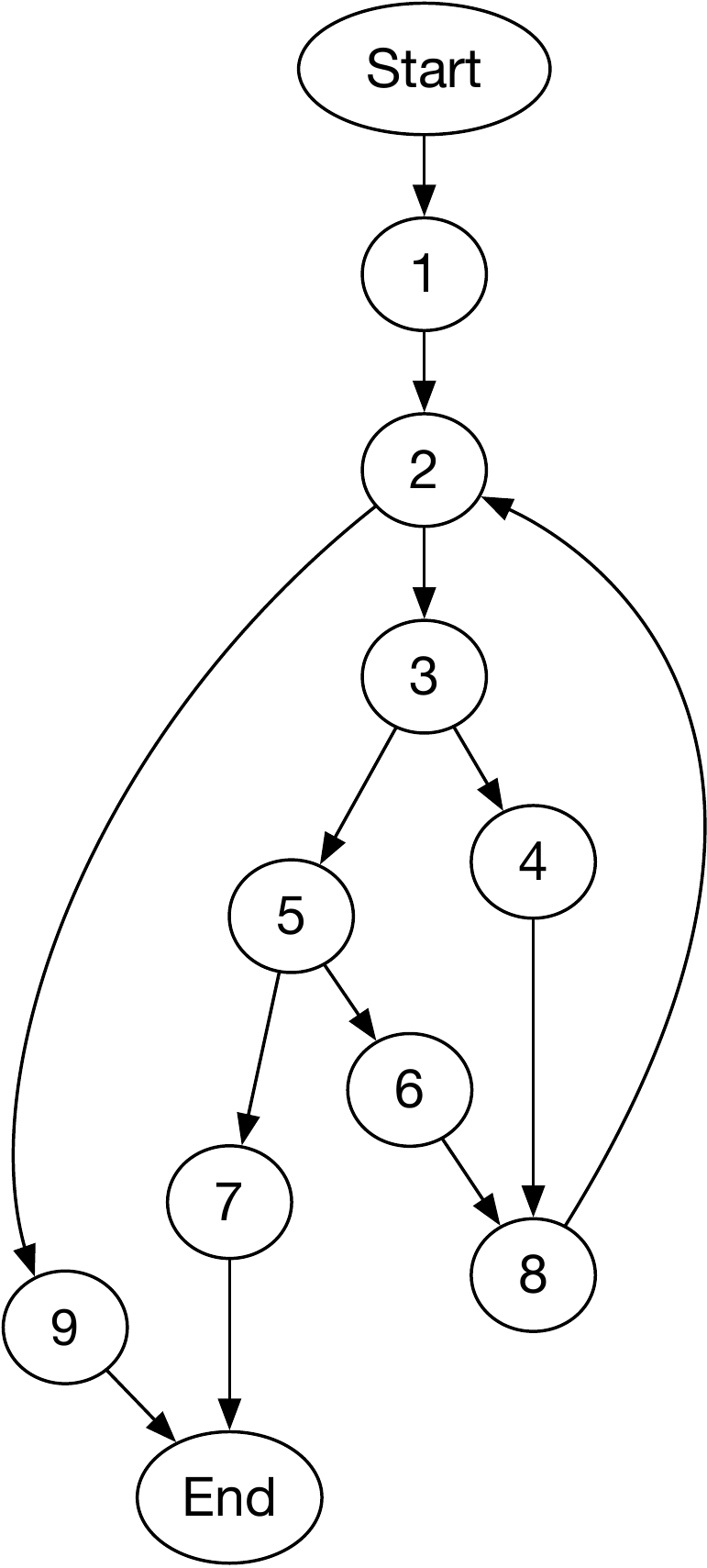}
 % \caption{java.util.Arrays.binarySearch0()}
  \label{fig:CFG1}
%\end{subfigure}
}
\caption{example method and corresponding CFG}
\label{fig:exampleCFG}
\end{figure}

This section presents some graph-theoretic  concepts that we utilize throughout this paper.  A  {\it directed graph} $G=(V, E)$ includes a set of vertices $V$ and a set of arcs $(v_i, v_j) \in E$, where $v_i, v_j \in V$. A {\it simple path} $p$ in $G$ is a sequence  of vertices $v_1, \cdots, v_k$, where each arc $(v_i, v_{i+1})$ belongs to $E$ for $1 \leq i < k$ and $k > 0$, and no vertex appears more than once in $p$ unless $v_1 = v_k$. A vertex $v_j$ is {\it reachable} from another vertex $v_i$ iff (if and only if) there is a simple path that emanates from $v_i$ and terminates at $v_j$. A SCC in $G$ is a sub-graph $G'=(V',E')$, where $V' \subseteq V$ and $E' \subseteq E$, and for any pair of vertices $v_i, v_j \in V'$, $v_i$ and $v_j$ are reachable from each other. Tarjan  \cite{tarjan1972depth} presents an algorithm with $O(|V|+|E|)$ time complexity that finds the SCCs of the input graph and constructs its {\it component graph}. Each vertex of the input graph appears in exactly one of the SCCs. The result is a Directed Acyclic Graph (DAG) whose every vertex is an SCC. A Control Flow Graph (CFG) models the flow of execution control between the basic blocks in a program, where a {\it basic block} is a collection of program statements without any conditional or unconditional jumps. A CFG is a directed graph, $G = (V, E)$. Each vertex $v \in V$ corresponds to a basic block. Each edge $e = (v_i, v_j) \in E$ corresponds to a possible transfer of control from block $v_i$ to block $v_j$. A CFG often has a {\it start vertex} that captures the block of statement starting with the first instruction of the program, and has some {\it end vertices} representing the blocks of statements that end in a halt/exit/return instruction. Figure \ref{fig:exampleCFG} illustrates an example method as well as its corresponding CFG (adopted from \cite{bang2015automatically}) for a class in the Apache Commons library. 

\begin{definition}[PP]
\label{def:pp}
A {\it PP} is a maximal simple path in a directed graph; i.e., a simple path that cannot be extended further without breaking its simplicity property (e.g., PP $\langle 2, 3, 4, 8, 2 \rangle$ in Figure \ref{fig:CFG1}). 
\end{definition}

\begin{definition}[CompletePP]
 \label{def:cmpltPP}
 	 A PP $p$ from $v_s$ to $v_t$ is a CompletePP {\it iff} $v_s$ is the Start vertex of $G$ and $v_t$ is an End vertex in $G$.  (e.g., the PP $\langle Start,1,2,3,5,7,End \rangle$ in Fig \ref{fig:CFG1})
 \end{definition}
 
\begin{figure}[!ht]
%\Figure[h]{
  \centering
\subfigure[Extracted SCC]{

  \includegraphics[scale= 0.4]{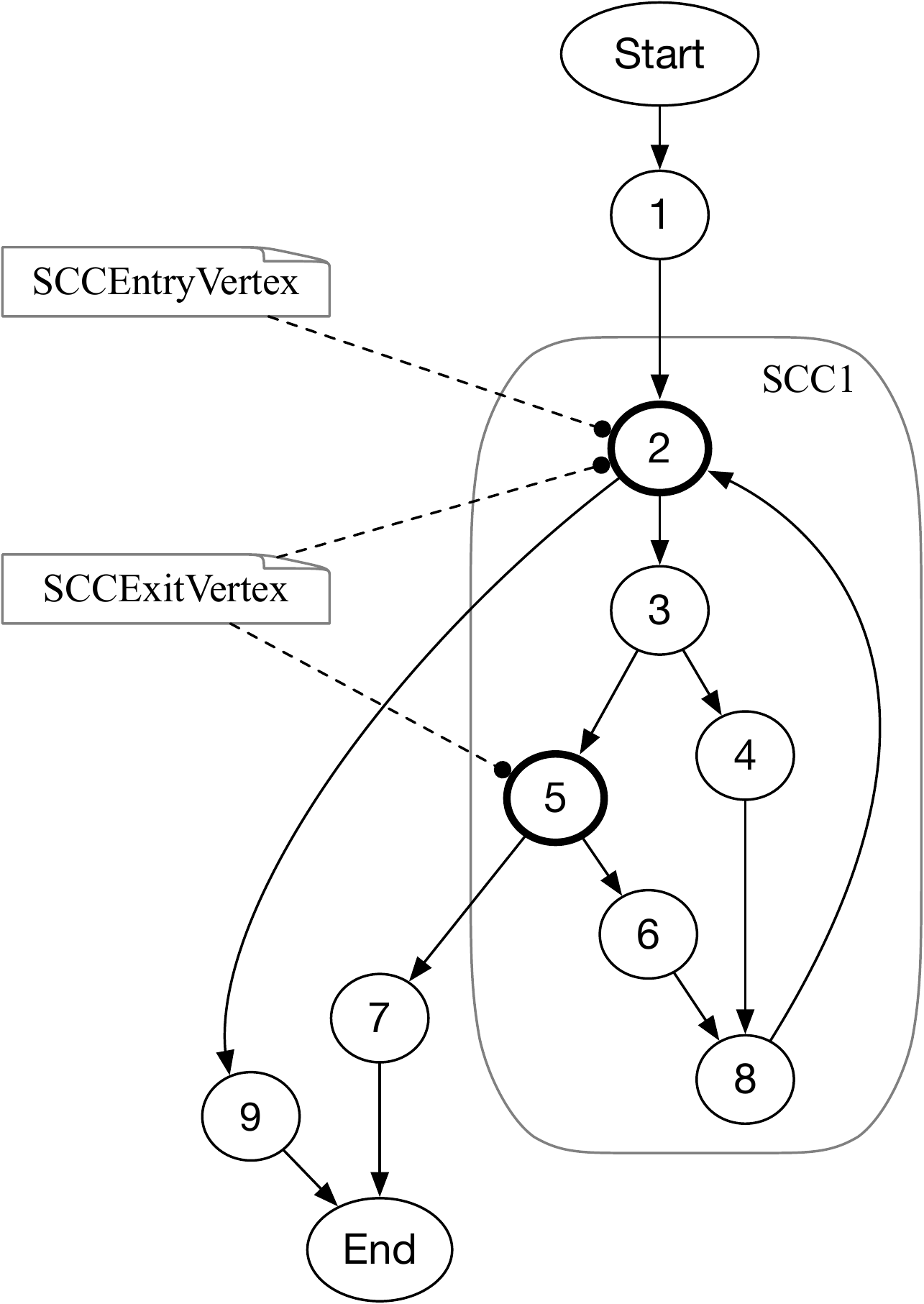}
%  \caption{java.util.Arrays.rangeCheck()}
  \label{fig:ExtractedSCC}
%\end{subfigure}%
}
\hspace*{3cm}
\subfigure[CCFG]{
  \centering
  \includegraphics[scale= 0.41]{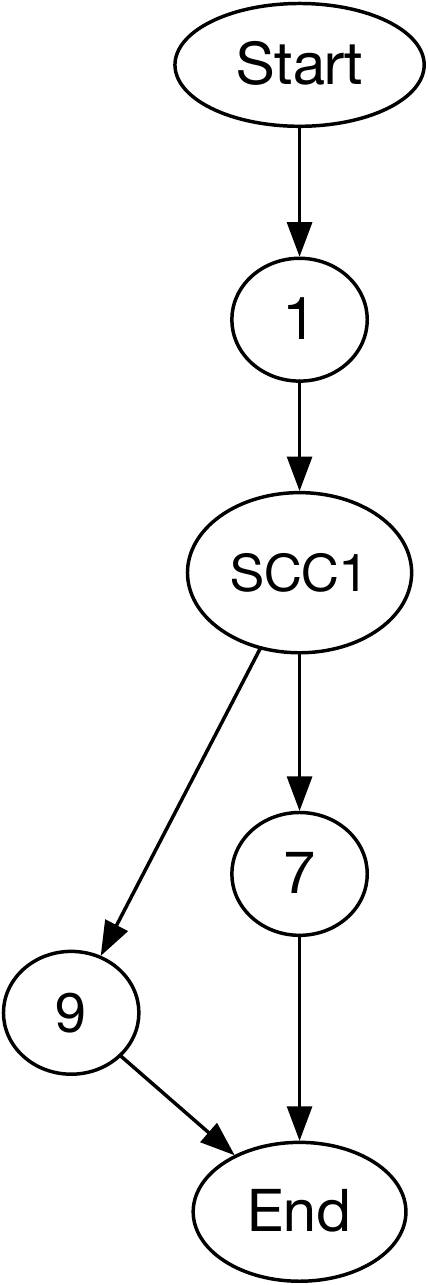}
 % \caption{java.util.Arrays.binarySearch0()}
  \label{fig:CCFG}
%\end{subfigure}
}
\caption{SCC and CCFG extracted from CFG for Fig \ref{fig:CFG1}}
\label{fig:exampleSCCandCCFG}
\end{figure}

\begin{definition}[Component Graph of CFGs]
\label{def:CCFG}
 	The component graph of a CFG $G=(V, E)$, called CCFG, is a DAG whose vertices are the SCCs of $G$, and its edges are the set of each arc $(v_i , v_j) \in E$ that starts in an $SCC_i$ and ends in $SCC_j$ (see Fig \ref{fig:CCFG}). 
 	\label{CCFG}
\end{definition}

In the following definitions, let $G=(V, E)$ be a CFG and  $C =(V_c, E_c)$ be an SCC in the CCFG of $G$; i.e., $C$ is a vertex in CCFG of $G$,

\begin{definition}[SccEntryVertex]
\label{def:SccEnV}
 A vertex $v_{en} \in V_c$ is an SccEntryVertex of $C$  {\it iff}  $~\exists v : v \in V \land v\not\in V_c : (v,v_{en}) \in E $. (e.g., Vertex 2 in Fig \ref{fig:ExtractedSCC}).
 \end{definition}
 
 \begin{definition}[SccExitVertex]
 \label{def:SccExV}
 A vertex $v_{ex} \in V_c$ is an  SccExitVertex of $C$  {\it iff}  $~\exists v : v \in V \land v\not\in V_c : (v_{ex},v) \in E $. (e.g., Vertices 2 and 5 in Fig \ref{fig:ExtractedSCC}).
 \end{definition}
 
 \begin{definition}[SccEntryExitPath]
 \label{def:SccEnExP}
An SccEntryExitPath is an acyclic simple  path from $v_{en} \in V_c$ to $v_{ex} \in V_c$, where $v_{en}$ is an SccEntryVertex and $v_{ex}$ is an SccExitVertex of $C$. (e.g., path $\langle 2 \rangle$ and $\langle 2,3,5 \rangle$ in Fig \ref{fig:ExtractedSCC}).
 \end{definition}

\begin{definition}[SccExitPath]
\label{def:SccExP}
 An {\it exit path} of $C$ is a simple path that starts in $v_s \in V_c$ and ends in $v_{ex} \in V_c$, where $v_s$ is not an SccEntryVertex but $v_{ex}$ is an SccExitVertex. We call $p$ an SccExitPath {\it iff} $p$ is a maximal exit path; i.e., $p$ is not a proper subpath of any other exit path in $C$. (e.g., path $\langle 3,5,6,8,2 \rangle$ in Fig \ref{fig:ExtractedSCC})
 \end{definition}

\begin{definition}[SccEntryPath]
\label{def:SccEnP}
  An {\it entry path} of $C$ is a simple path that starts in $v_{en} \in V_c$ and ends in $v_t \in V_c$, where $v_{en}$ is  an SccEntryVertex but $v_t$ is not an SccExitVertex. We call $p$ an SccEntryPath {\it iff} $p$ is a maximal entry path; i.e., $p$ is not a proper subpath of any other entry path in $C$. (e.g., the path $\langle 2,3,4,8 \rangle$ in Fig \ref{fig:ExtractedSCC})
 \end{definition}
 
 \begin{definition}[SccInternalPP]
 \label{def:SccIntPP}
An SccInternalPP $p$ of $C$ is a PP starting in $v_s \in V_c$ and ending at $v_t \in V_c$.  Moreover, if $p$ is  acyclic, then  it must not start at an SccEntryVertex in $C$ or terminate in some SccExitVertex in $C$. (e.g., the PP $\langle 3,4,8,2,3 \rangle$ in Fig \ref{fig:ExtractedSCC})
 \end{definition}
 
 \begin{definition}[SccExitPP]
 \label{def:SccExitPP}
A PP $p$ from  $v_s$ to $v_t$ is an SccExitPP {\it iff} $v_s\in V_c$ and $v_t \in V$ is an End vertex of $G$. (e.g., the PP $\langle 3,5,6,8,2,9,End \rangle$ in Fig \ref{fig:ExtractedSCC})
 \end{definition}
 
 \begin{definition}[SccEntryPP]
 \label{def:SccEnPP}
 A PP $p$ from $v_s$ to $v_t$ is an SccEntryPP  {\it iff} $v_s$ is the Start vertex of $G$ and  $v_t \in V_c$. (e.g., the PP $\langle Start,1,2,3,4,8 \rangle$ in Fig \ref{fig:ExtractedSCC})
 \end{definition}

\subsection{Compositional Method for PPs and TPs Generation}

\begin{figure}[t]
\vspace*{5mm}
\centering
%\hspace*{-1mm}
\includegraphics[scale = .45]{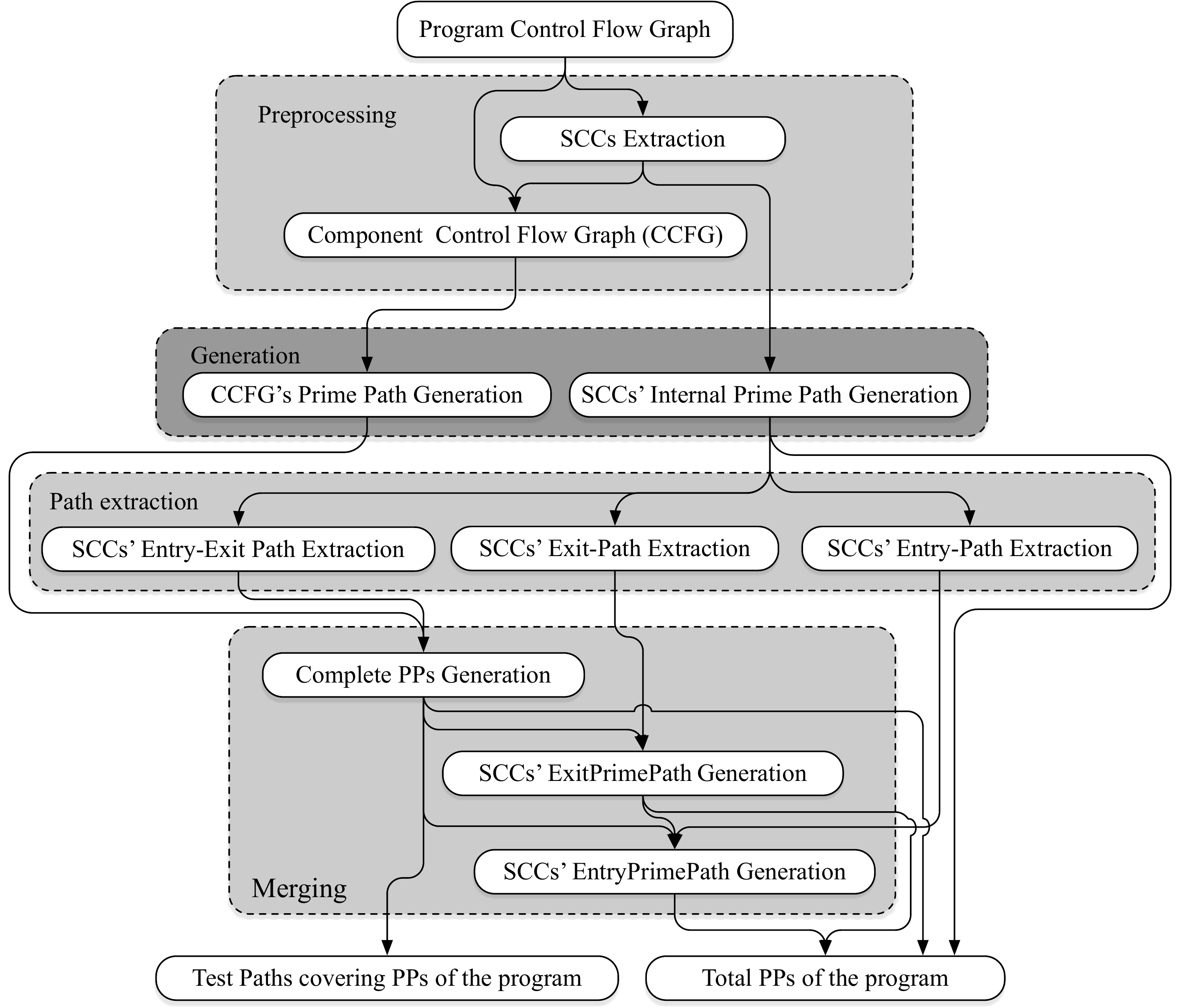}
%\epsfig{file = fig/structure1.eps,width=10.3cm,height=9.3cm}

%\vspace*{-63mm}
\caption{Overview of the Compositional Method \cite{fazli2019time}}
%\vspace*{-1mm}
\label{fig:ParallCompMethod}
\end{figure}

This section represents an overview of the CPU-based compositional method introduced in \cite{fazli2019time} that provides better time efficiency in comparison with existing methods for PPs generation. The basic idea behind such a compositional method is to (1) compute the component graph of the input CFG, denoted CCFG; (2) generate the set of PPs of CCFG and the set of PPs of each individual SCC in CCFG; (3) extract different types of intermediate paths related to each SCC, and (4) merge the PPs of SCCs to generate all PPs of the original input CFG. The proposed method distinguishes four types of PPs defined in Definitions \ref{def:cmpltPP}, \ref{def:SccIntPP}, \ref{def:SccExitPP} and \ref{def:SccEnPP}. Figure \ref{fig:ParallCompMethod} illustrates the steps of the SCC-based compositional method as well as different types of paths we generate.

%\subsubsection{Preprocessing}
%\label{preprocessing}

\noindent{\bf Preprocessing}.\ In this phase we first identify the set of all SCCs of the input CFG G = (V , A) , and then construct the Component Control Flow Graph (CCFG).

%\subsubsection{Generation}
%\label{generation}

\noindent{\bf PP Generation}.\
This phase separately generates PPs of all extracted SCCs and the constructed CCFG using the vertex-based algorithm described in \cite{fazli2019time}. These tasks are accomplished by applying the proposed algorithm of \cite{fazli2019time} on each SCC with more than one vertex. For example, running the vertex-based algorithm with the CCFG in Figure \ref{fig:CCFG} as input yields two PPs $\langle Start,1,SCC_1,9,End\rangle$ and $\langle Start,1,SCC_1,7,End\rangle$. Moreover, applying this algorithm on the SCC of Figure \ref{fig:ExtractedSCC}, generates 11 internal PPs: $\langle 4,8,2,3,5,6\rangle$, $\langle 3,5,6,8,2,3\rangle$, $\langle 2,3,5,6,8,2\rangle$, $\langle 5,6,8,2,3,4\rangle$, $\langle 8,2,3,5,6,8\rangle$, $\langle 6,8,2,3,5,6\rangle$, $\langle5,6,8,2,3,5 \rangle$, $\langle 4,8,2,3,4\rangle$, $\langle 3,4,8,2,3\rangle$, $\langle 2,3,4,8,2\rangle$, $\langle 8,2,3,4,8\rangle$. According to the definition of SccInternalPPs, some of the PPs generated for each SCC and CCFG in this phase do not necessarily belong to the main CFG's PPs (e.g., acyclic SccInternalPPs starting from a SccEntryVertex or terminating at a SccExitVertex). These types of paths are sub paths of the main CFG's PPs (which are produced in subsequent phases).

%\subsubsection{Path extraction}
%\label{pathExtraction}

\noindent{\bf Path extraction}.\
In this phase, we extract SccEntryExitPaths (Definition \ref{def:SccEnExP}), SccExitPaths (Definition \ref{def:SccExP}) and SccEntryPaths (Definition \ref{def:SccEnP}) leveraging corresponding algorithms. The extracted paths are sub paths of the main CFG's PPs crossing through some SCCs. 
The SccEntryExitPaths extraction algorithm \cite{fazli2019time} receives the internal PPs of a given SCC, a specified entry vertex $v_{en}$ and a specified exit vertex $v_{ex}$. Then, this algorithm  chooses the longest distinct sub paths starting from $v_{en}$ and ending at $v_{ex}$. For example, if the SccEntryExitPaths extraction algorithm takes those eleven internal PPs of the SCC of Figure \ref{fig:ExtractedSCC}, along with the entry vertex $\{2\}$ and the exit vertices $\{2,5\}$ as inputs, it extracts $\langle2\rangle$ and $\langle2,3,5\rangle$ as that SCC's entry exit paths. The first part of each SccExitPPs are SccExitPaths included in a SCC. We use  SccExitPaths extraction algorithm \cite{fazli2019time} to extract all SccExitPaths. For all internal PPs of a given SCC, this algorithm reaps all distinguished sub paths starting from first to a specified exit vertex $v_{ex}$. The paths $\langle3,5,6,8,2\rangle$, $\langle3,4,8,2\rangle$, $\langle4,8,2,3,5\rangle$, $\langle6,8,2,3,5\rangle$ are the output of applying this algorithm on the SCC of Figure \ref{fig:ExtractedSCC}, and the exit vertices $\{2,5\}$.

The fourth type of the main CFG's PPs are SccEntryPPs finishing at a given SCC and beginning from the start vertex or another SCCs. The last part of this PPs (SccEntryPaths) belongs to the given SCC. We employ the SccEntryPaths extraction algorithm \cite{fazli2019time} to extract all SccEntryPaths in a given SCC. For all internal PPs, this algorithm selects all different longest sub paths starting from a given entry vertex $v_{en}$ to the end. Applying the algorithm on the SCC of Figure \ref{fig:ExtractedSCC} and the entry vertex $\{2\}$ results in $\langle2,3,4,8\rangle$ and $\langle2,3,5,6,8\rangle$ as output.

%\subsubsection{Merging}
%\label{merging}

\noindent{\bf Merging}.\
 The compositional method merges the PPs generated in the previous phase towards generating PPs and TPs. The inputs of this phase include the produced PPs of CCFG and the extracted SccEntryExit, SccExit and SccEntry paths from path extraction phase. All the PPs of the input CFG that pass over some SCCs are produced in this phase. The outputs of the merging phase, therefore,  are all CompletePPs, SccExitPPs and SccEntryPPs of the input CFG according to corresponding algorithms. 

The method in \cite{fazli2019time} includes an algorithm  that generates all Complete PPs. The input of this algorithm consists of PPs of the CCFG along with SccEntryExit paths of all SCCs. The output of this algorithm includes all complete PPs of the input CFG crossing through some SCCs. For each PP $p$ of CCFG including any $SCC_i$, this algorithm replaces $SCC_i$ with those SccEntryExitPaths of the $SCC_i$ that can be integrated with $p$. For example the complete path of CCFG $\langle Start,1,SCC_1,7,End\rangle$ of Figure \ref{fig:exampleSCCandCCFG} can tour the $SCC_1$'s entry exit path $\langle2,3,5\rangle$ since there are edges $\{\langle1,2\rangle,\langle5,7\rangle \in E\}$ in the main CFG. These substitutions create new complete PPs of the main CFG. The consequence of running the CompletePPs generator algorithm on the CCFG's PPs $\langle Start,1,SCC_1,9,End\rangle$, $\langle Start,1,SCC_1,7,End\rangle$ with SccEntryExit paths $\langle2\rangle$, $\langle2,3,5\rangle$ of Figure \ref{fig:exampleSCCandCCFG} are two complete PPs
 $\langle Start,1,2,9,End \rangle$, $\langle Start,1,2,3,5,7,End \rangle$ of the input CFG.

The merging phase also extracts all SCC's exit PPs. This task is accomplished by the SccExitPaths generator algorithm of \cite{fazli2019time}. This algorithm produces all PPs exited from a SCC and finished at an $End$ vertex. Assuming the complete PP $p$ leaves $SCC_i$ through an exit vertex $v_{ex}$, the algorithm selects a subpath called $r$, from $v_{ex}$ to the end vertex of $p$. Then, the algorithm merges all $SCC_i$'s exit paths terminating at $v_{ex}$ with the $r$. The result of applying the SccExitPaths generator algorithm \cite{fazli2019time} on the CFG's complete PPs $\langle Start,1,2,9,End \rangle$, $\langle Start,1,2,3,5,7,End \rangle$ with SCC Exit paths $\langle3,5,6,8,2\rangle$, $\langle3,4,8,2\rangle$, $\langle4,8,2,3,5\rangle$, $\langle6,8,2,3,5\rangle$  of Figure \ref{fig:exampleSCCandCCFG} are four SCC's exit PPs
 $\langle3,5,6,8,2,9,End \rangle$, $\langle 3,4,8,2,9,End \rangle$, $\langle 4,8,2,3,5,7,End \rangle$ and $\langle 6,8,2,3,5,7,End \rangle$ for the input CFG.

The fourth type of PPs, SccEntryPPs contain those PPs that emanate from the start vertex or another SCC and end in a given SCC. The SccEntryPPs generator algorithm \cite{fazli2019time} generates all SccEntryPPs to each SCC. The inputs of this algorithm is the SccEntryPaths, the CompletePPs, and the SccExitPPs. For each CompletePP or SccExitPP $p$ that contains the $SCC_i$'s entry vertex $v_{en}$, the algorithm selects a subpath from the beginning to $v_{en}$ called $r$. Then, this algorithm concatenates all $SCC_i$'s Entry Paths started by $v_{en}$ with $r$. The output of running the SccEntryPPs generator algorithm \cite{fazli2019time} on the CFG's complete PPs $\langle Start,1,2,9,End \rangle$, $\langle Start,1,2,3,5,7,End \rangle$ with SCC entry paths $\langle2,3,4,8\rangle$, $\langle2,3,5,6,8\rangle$ of Figure \ref{fig:exampleSCCandCCFG} are two SCC's entry PPs $\langle Start,1,2,3,4,8\rangle$, $\langle Start1,2,3,5,6,8\rangle$ of the main CFG. Finally, by applying the compositional method for PPs and TPs Generation on the CFG Figure \ref{fig:CFG1}, we reach 19 PPs of all four types: 11 SCC internal PPs, 2 complete PPs, 4 SCC exit PPs, and 2 SCC entry PPs.

\section{Problem Statement}
\label{sec:ProblemStatement}
\begin{figure}[h]
%\Figure[h]{
  \centering
\subfigure[\scriptsize CFG with out-degree more than 2]{

  \includegraphics[width=4.7cm,height=2.1cm]{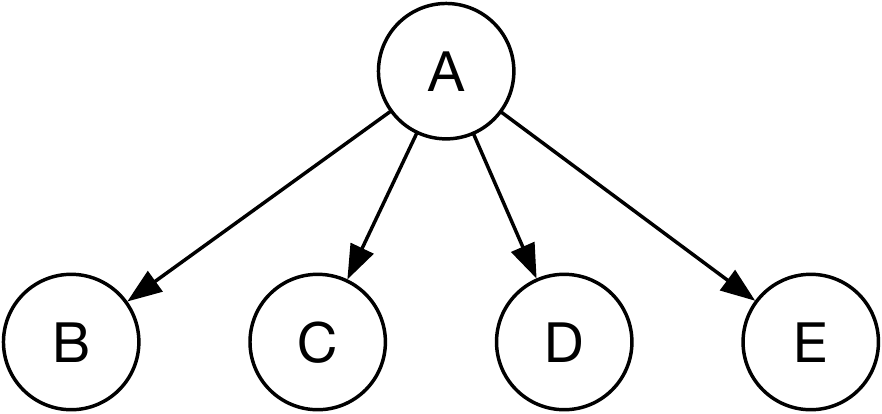}
%  \caption{java.util.Arrays.rangeCheck()}
  \label{fig:sourceGraph}
%\end{subfigure}%
}
\hspace*{1.5cm}
\subfigure[\scriptsize Converted CFG with out-degree 2]{
  %\centering
  \includegraphics[width=4.7cm,height=3.1cm]{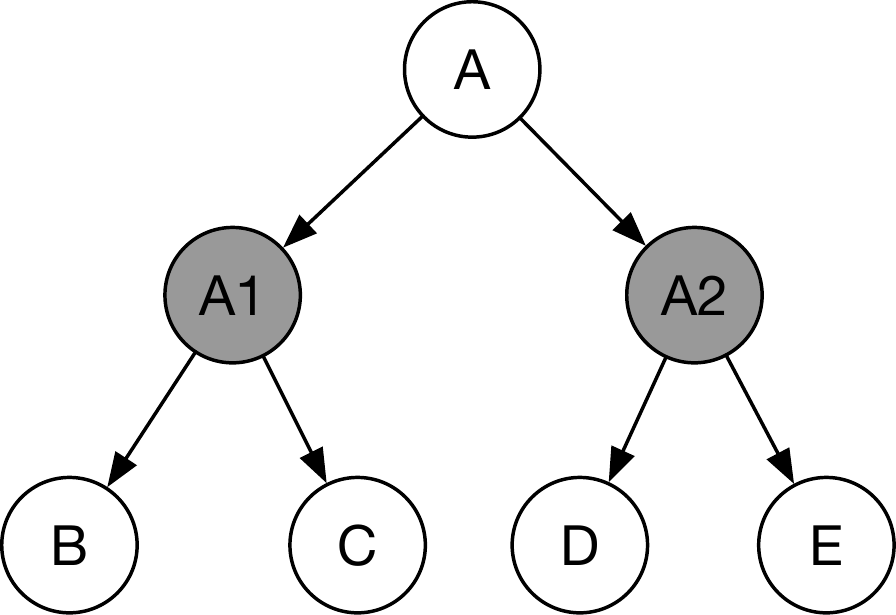}
 % \caption{java.util.Arrays.binarySearch0()}
  \label{fig:convertedGraph}
%\end{subfigure}
}
\caption{Generating CFGs with maximum out-degree 2}
\label{fig:GeneratingOutDeg2}
\end{figure}

Generating PPs of the control flow graphs related to real world programs with a large Npath complexity is an important problem in software structural testing. These types of graphs have a huge number of PPs and processing them under conventional algorithms on CPUs requires a lot of time. Therefore, it is necessary to develop algorithms that, eliminate this weakness and maintain the accuracy of the PP generation while significantly reducing the processing time of the mentioned graphs. In a graph-theoretic setting, the PPs generation problem can be formulated as follows:

\begin{problem}[\bf PPs Generation]
\label{probstmt}
\
\begin{itemize}
\item {\bf Input}: A graph $G=(V,E)$ that represents the CFG of a given program,  a start vertex $s \in V$ and an end vertex $e \in V$.
 
 \item {\bf Output}: The set of PPs finished at each vertex $v \in V$ and the set of TPs covering all PPs.
 \end{itemize}
 \end{problem}
We note that the complexity of solving Problem is higher when the input graph is an SCC because every vertex is reachable from any other vertex in an SCC, and this drastically increases the number of PPs. For this reason, next section proposes a parallel GPU-based algorithm that extracts the PPs of SCCs in a time and space efficient fashion. The in-degree of $s$ is 0, and out-degree of $e$ is 0. We focus on CFGs where all vertices $v \in V$ except $e$ have a maximum out-degree of 2. Without loss of generality, we can convert a vertex $v$ with an out-degree greater than 2 (i.e, switch case structure) to a vertex with out-degree 2 by adding some new intermediate vertices between $v$ and its successor vertices (see Figure \ref{fig:GeneratingOutDeg2}).

\section{Data Structures}
\label{sec:dataStruct}
In order to scale up PPs generation, this section presents a parallel compositional method that provides better time efficiency in comparison with existing sequential methods for PPs generation. This section describes a GPU-based  PP generation algorithm, the data structure for storing the input CFG, path data structure, and memory allocated to all generated PPs.
The input of the PP generation algorithm (Algorithm \ref{alg:ParallelDevice}) includes a CFG representing the Program Under Test (PUT). The output of Algorithm \ref{alg:ParallelDevice} is the set of all PPs finished at each vertex $v_i \in V$. 

\subsection{CFG Data Structure}
\begin{figure}[t]
\vspace*{1mm}
\centering
%\hspace*{-1mm}

\includegraphics[scale= 0.5]{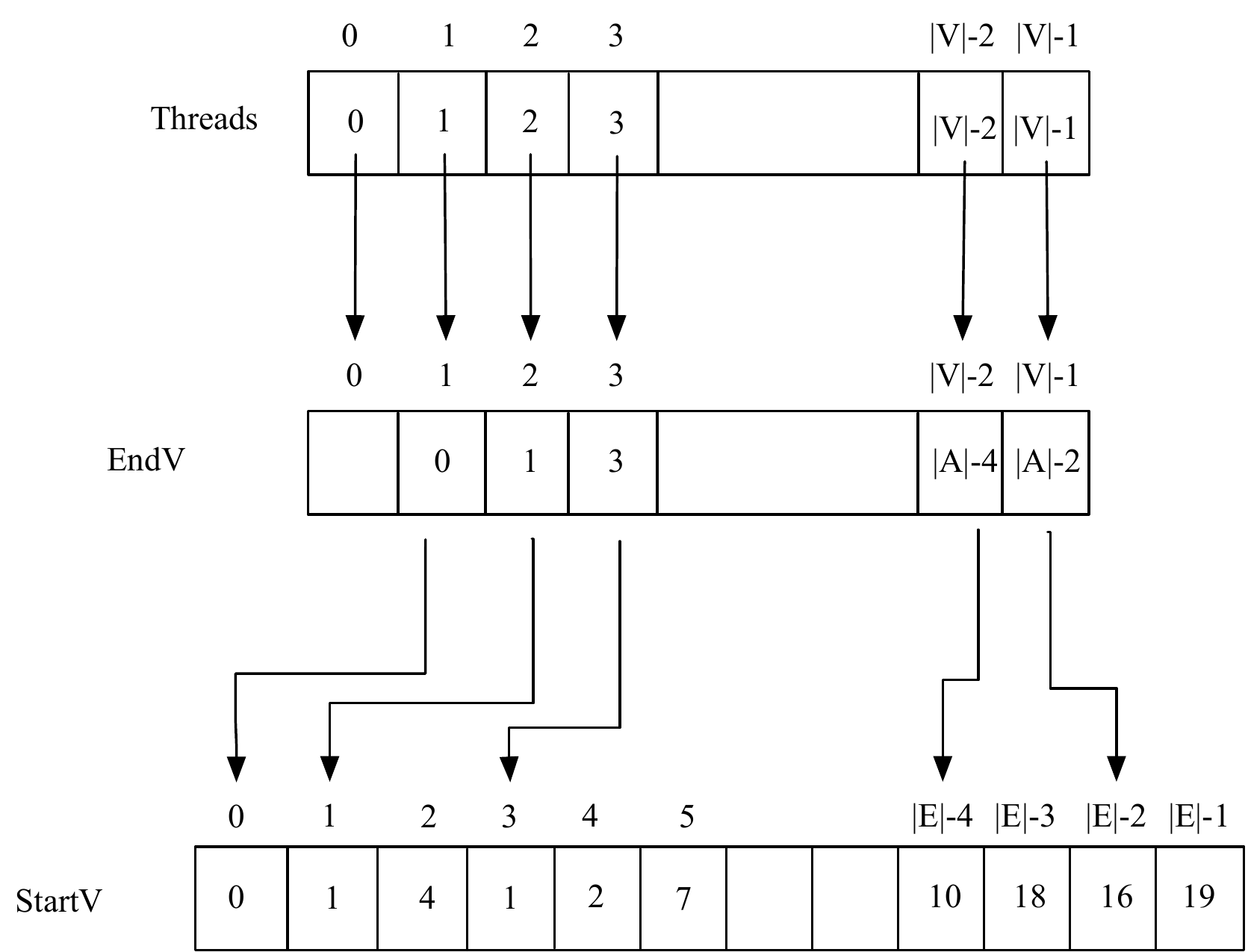}
%\epsfig{file = fig/structure1.eps,width=10.3cm,height=9.3cm}

%\vspace*{-63mm}
\caption{Compressed Sparse Row (CSR)  graph representation.}
%\vspace*{-1mm}
\label{fig:MemDesign}
\end{figure}

A matrix is usually stored as a two-dimensional array in memory. In the case of a sparse matrix, memory requirements can be significantly reduced by maintaining only non-zero entries. Depending on the number and distribution of non-zero entries, we can use different data structures. The Compressed Sparse Row (CSR, CRS or Yale format) \cite{eisenstat1981algorithms} represents a matrix by a one-dimensional array that supports efficient access and matrix operations. We employ the CSR data structure (see Figure \ref{fig:MemDesign}) to maintain a directed graph in the global memory of GPUs, where vertices of the graph receive unique IDs in $\{0, 1, \cdots, |V|-1\}$. To represent a graph in CSR format,  we store end vertices and start vertices of arcs in two separate arrays $EndV$ and $StartV$ respectively (see Figure \ref{fig:MemDesign}). Each entry in EndV points to the staring index of its adjacency list in array StartV. We assign one thread to each vertex. That is, thread $t$ is responsible for the vertex whose ID is stored in $EndV [t]$, where $\{0 \leq t < |V |-1\}$ (see Figure \ref{fig:MemDesign}). Since the proposed algorithm computes all PPs ending in each vertex $v \in V$, maintaining the predecessor vertices is of particular importance. In CSR data structure, first the vertex itself and then its predecessor vertices are stored.

\subsection {Path Structure}
\begin{figure}[t]
\vspace*{1mm}
\centering
%\hspace*{-1mm}

\includegraphics[scale= 0.5]{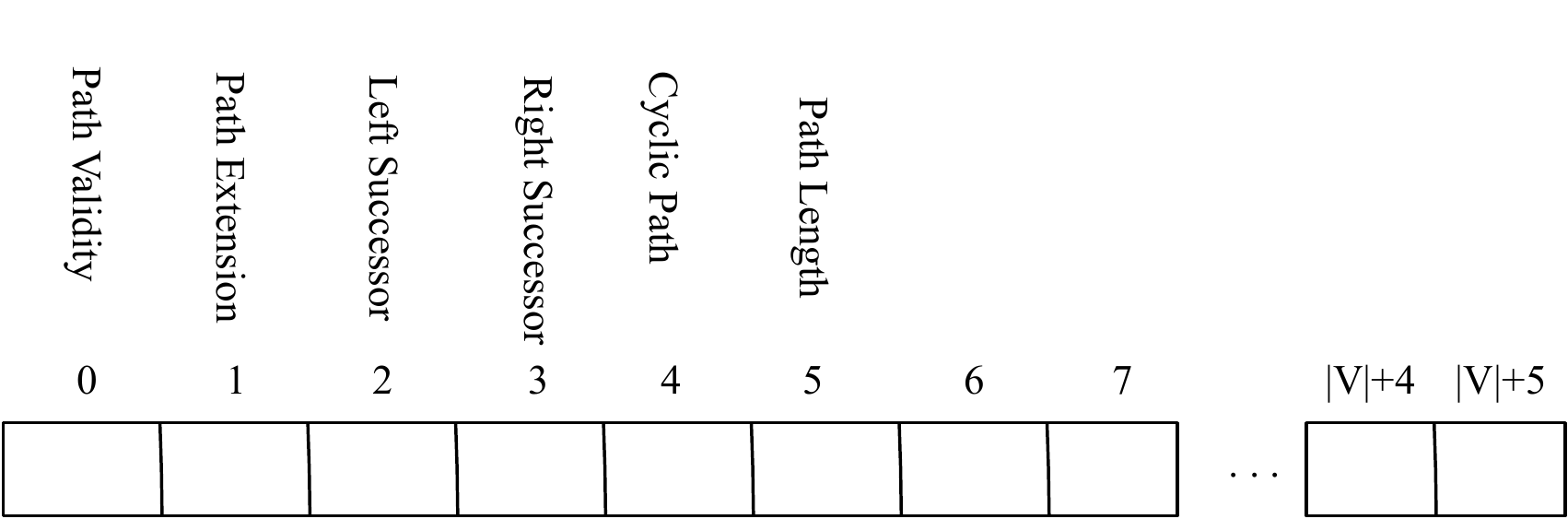}
%\epsfig{file = fig/structure1.eps,width=10.3cm,height=9.3cm}

%\vspace*{-63mm}
\caption{Path structure}
%\vspace*{-1mm}
\label{fig:Pathstructure}
\end{figure}

We utilize a set of flags to keep the status of each recorded path along with each vertex (see Figure \ref{fig:Pathstructure} ). Let $v_i$ be a vertex and $p$ be a path associated with $v_i$. The PathValidity flag ($p[0]$) indicates whether or not the recorded information represents a simple path. The PathExtension flag ($p[1]$) means that the current path is an extended path; hence not a PP. We assume each non-final vertex can have a maximum of two successor vertices. We use the LeftSuccessor ($p[2]$) and the RightSuccessor ($p[3]$) flags to indicate whether the thread of each corresponding successor has read the path ending in vertex $v_i$. Once one of those successor threads reads the path ending in $v_i$ it will mark its flag. In each iteration of the algorithm, paths with marked extension and marked successor flags will be pruned. We set the CyclicPath flag ($p[4]$) if $p$ is a cyclic path. If $p$ is cyclic, then it will no  longer be processed by the successor threads of $v$ and is recorded as a PP at the $v_i$. (see Figure \ref{fig:Pathstructure}).

{\bf Note}: Since there is a unique thread associate to each vertex, we use the terms "successor" and "predecessor" for both vertices and threads.

\subsection{ Memory Allocation} 
\begin{figure}[h]
\vspace*{-5mm}
\centering
%\hspace*{-1mm}
\includegraphics[scale= 0.5]{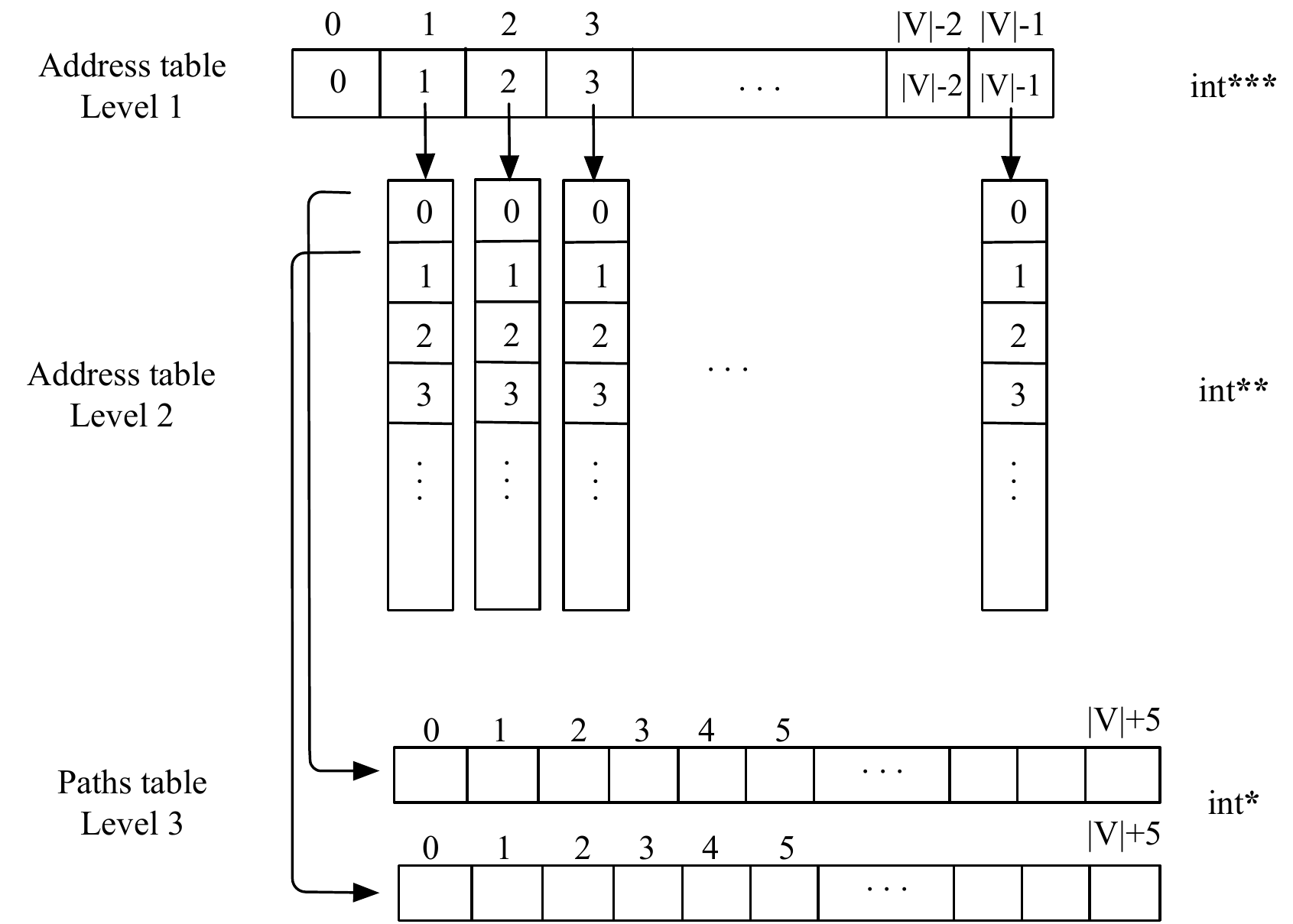}
%\epsfig{file = fig/structure1.eps,width=10.3cm,height=9.3cm}

%\vspace*{-63mm}
\caption{Three-level Path Accessing Method}
%\vspace*{-1mm}
\label{fig:PathsDesign}
\end{figure}

In each CFG, the extraction of the PPs is based on the generation of the simple paths terminated in each vertex $v_i \in V$. There is a list associated to each vertex $v_i$, denoted $v_i.list$ to record all generated PPs ending in $v_i$. To implement this idea, all acyclic paths ending in predecessors of $v_i$ must be copied to the list of the vertex $v_i$. For a large CFG, the number of such paths could be very large, which would incur a significant space cost on the algorithm. To mitigate this space complexity, we introduce a non-contiguous memory allocation method with a pointer based Three-level Path Accessing Method (TPAM). TPAM is a path accessing scheme which consists of three levels of address tables in a hierarchical manner. The entries of Level 1 address table with length $|V|$ are pointers to each $v_i.list$ at Level 2 address tables. Level 2 address tables contain addresses of all paths stored in each $v_i.list$. The entries of the last level tables are actual paths information in the memory (see Figure \ref{fig:PathsDesign}).

All activities such as compare, copy, extend and delete are applied on the paths of each vertex. Let $v_i$ be a vertex in $V$ and $p$ be a path in $v_i.list$. To access path $p$, the start address of the $v_i.list$ is discovered from the first array (i.e, $Path[v_i]$). The start address of path $p$ is stored in Table 2 (i.e, $Path[v_i][p]$). The list of vertices of path $p$ is in Table 3, which according to path structure mentioned in Figure \ref{fig:Pathstructure}, all activities can be done on the elements of the path (i.e, $Path[v_i][p][5]$ shows the length of the path $p$).

Instead of using malloc in allocating host memory, we call CUDA to create page-locked pinned host memory. Page-Locked Host Memory for CUDA (and other external hardware with DMA capability) is allocated on the physical memory of the host computer. This allocation is labeled as non-swappable (not-pageable) and non-transferable (locked, pinned). This memory can be accessed with the virtual address space of the kernel (device). This memory is also added to the virtual address space of the user process to allow the process to access it. Since the memory is directly accessible by the device (i.e., the GPU), the write and read speeds are high bandwidth. Excessive allocation of such memory can greatly reduce system performance as it reduces the amount of memory available for paging, but proper use of this memory allocation method provides a high performance data transfer schema.

%\vspace*{-3mm}
\section{GPU-based PP Generation Algorithm}
\label{sec:parAlg}
\vspace*{-2mm}

A GPU-based CUDA program has a CPU part and a GPU part. The CPU part is called the host and the GPU part is called the kernel, capturing an array of threads. The algorithm proposed in this section includes one kernel (illustrated in Algorithm \ref{alg:DeviceKernel}). The host (i.e., CPU) (Algorithm \ref{alg:Host}) initializes the $v_i.list$ of all $v_i$ and an array of boolean flags, called $PublicFlag$, where $PublicFlag[v_i] = true$ indicates that the predecessors of vertex $v_i$ have been updated and so the $v_i.list$ needs to be updated. One important objective is to design a self-stabilizing algorithm with no CPU-GPU communications, thus the host launches Cuda-Kernel-UpdateVertex only once. The proposed algorithm is implemented in such a way that there is no need for repeated calls to synchronize different threads in the block. One of the major challenges in parallel applications that drastically reduces the efficiency of these types of implementations is the use of atomic instructions. These types of instructions are executed completely without any interruption, but greatly reduce the efficiency of parallel processing. The self-stabilizing device (i.e., GPU) code in this section is implemented without using atomic instructions.

%\vspace*{-1cm}

\begin{algorithm}[H]
\caption{Device: Cuda-Kernel-Update-Vertex}
\begin{algorithmic} [1]
	\For{\textbf{each} thread assigned to vertex $v_i$}
	\State Update-Vertex($v_i$);
	\EndFor
\end{algorithmic}
\label{alg:DeviceKernel}
\end{algorithm}

\begin{algorithm}[H]
\caption{Host: Generating the set of all PPs of the input CFG}
\hspace * {\algorithmicindent} \textbf{Input:} $G(V,A)$ with an out-degree \ 2, Start vertex $s \in V$\\
\hspace * {\algorithmicindent} \textbf{Output:} The set of PPs ending in each vertex $v \in V$ \\
\hspace * {\algorithmicindent} \textbf{Initialize:} 
$\forall v_i \in V$, $v_i.list = \{v_i\}$, $v_i.PublicFlag = true$.
\begin{algorithmic} [1]
	\State Cuda-Kernel-UpdateVertex ;
\end{algorithmic}
\label{alg:Host}
\end{algorithm}

\begin{algorithm}[H]
\caption{Update-Vertex($v_i$)}
\begin{algorithmic} [1]
%	\While {$\exists v_{i} \in V,$ where $ v_i $ is unmarked}
	\While {($v_i.PublicFlag=true$)}
	%or ($\exists v: v \in Q: v$ is not marked)
	   % if $v_i$ is not marked.
		\For{\textbf{each} path $  p \in v_{i}.list $}
			\If { $p$ has been read by both successors } 
				\If {($p$ is an extended path) or ($p$ is already included in some path $p' \in v_{i}.list$)}
					\State remove $p$;
	%			\Else {\textbf{if }$p$ is covered by any $p' \in v_{i}.list$ \textbf{then}}
%					\State remove $p$;
				\EndIf

			\EndIf

		\EndFor
	%	\State $v_i.LocalFlag = false$;
		\For{\textbf{each}  $ v_{j} $ where $(v_{j},v_{i})\in A $ }             // Read from predecessors.
			\For{\textbf{each} path $ q \in v_{j}.list $ } // PathValidity flag
				\If{$q$ is not read by $v_i$}
				    \State Label $q$ as read by $v_i$;  // Left or Right Successor flag
					\If {($q$ is not a cycle) \textbf{and} ($v_{i}$  does not appear in $q$  \textbf{or} $v_i$ is the first vertex of $q$)}
						\State ExtendPath ($q$,$v_{i}$);
						\State Label $q$ as an extended path;   // PathExtension flag
			%			\State $v_i.LocalFlag = true$; 
					
				   \EndIf
				\EndIf
			\EndFor
		\EndFor
		% and mark $v_i$.
		
		\State $v_i.LocalFlag = false$;
		\For{\textbf{each} path $  p \in v_{i}.list $}
			\If {$p$ is not read by both successors} 
					\State $v_i.LocalFlag = true$;
	%			\Else {\textbf{if }$p$ is covered by any $p' \in v_{i}.list$ \textbf{then}}
%					\State remove $p$;
				\EndIf

		\EndFor
		\If {$v_i.LocalFlag = false$}   // all paths in $v_i.list$ have been read by both $v_i's$ successors 
			\State $v_i.PublicFlag = false$;
			\For{\textbf{each}  $ v_{k} $ where $v_{i}$ is Reachable from $v_{k}$}
				\If{$v_k.PublicFlag = true$}
					\State $v_i.PublicFlag = true$;
				\EndIf
				% as an unmarked vertex.
		\EndFor
		\EndIf
	\EndWhile
	\end{algorithmic}
\label{alg:ParallelDevice}
\end{algorithm}

  Algorithm \ref{alg:ParallelDevice} forms the core of the kernel Algorithm \ref{alg:DeviceKernel}. This Algorithm performs three kinds of processing on each vertex $v_i \in V$: pruning the extended paths in $v_i.list$, extending  acyclic simple paths in the lists of all predecessor vertices of $v_i$, and examining the termination of all backward reachable vertices of the $v_i$.  {\bf Lines 2 to 8} in Algorithm \ref{alg:ParallelDevice} remove   all extra paths from $v_i.list$. The path $P \in v_i.list$ is an extra path if it has been extended by one of the $v_i$'s successor(s) or covered by another path $P' \in v_i.list$.   {\bf Lines 9 to 19} extend  eligible acyclic simple paths in the lists of all predecessor vertices of $v_i$.  Suppose that $v_j \in V$ is one of the $v_i$'s predecessor. A path $q \in v_j.list$ is an eligible path if $q$ is not a cyclic path and if $v_i$ appeared in $q$ then $v_i$ is the start vertex of $q$. The thread assigned to $v_i$ runs a function called {\it ExtendPath} (in Algorithm \ref{Function:ExtendPath}) to append the new eligible path to the $v_i.list$.  In {\bf -Lines 21 to 33},  the thread  of $v_i$ cannot be terminated if the vertex $v_i$ is not the final vertex and has any unread paths in $v_i.list$ (Lines 21 to 25).  Thread  of $v_i$ then examines the termination of all its backward reachable vertices by examining their $PublicFlag$. If all the previous vertices of the $v_i$ are terminated, then the vertex $v_i$ will also set its {\it PublicFlag} to   false and exit the while loop (Lines 28 to 32).

  We devise Algorithm \ref{Function:ExtendPath} to append a new simple path to the list of a given vertex. This algorithm takes a path $p$ as well as a specified vertex $v$ as inputs. Algorithm \ref{Function:ExtendPath} first adds the vertex $v$ at the end of the path $p$ and increments the length of $p$ (Lines 2 and 3). Then, it checks the occurrence of vertex $v$ as the first vertex of $p$. This property causes the new path $p$ to be considered as a cycle in vertex $v$ (Lines 4 to 6). Finally, Algorithm \ref{Function:ExtendPath} sets PathValidity flag of the new path $p$ to true and appends it to the end of $v.list$ (Lines 7 and 8).

\subsection{Proof of Termination}
This section proves that TPGen eventually terminates.

\begin{lemma}
\label{readOnce}
Let $(v_j, v_i)$ be an arc in $A$ and $q$ be a path in $v_j.list$. The condition of the if statement on Line 11 in Algorithm \ref{alg:ParallelDevice} evaluates to true for $q$ and $v_i$ no more than once. That is, each path $q \in v_j.list$ is read by $v_i$ at most once.
\end{lemma}

\begin{proof}
Suppose Algorithm \ref{alg:ParallelDevice} has already entered the if statement on Line 11 for some $q$ and $v_i$. This means that $q$ has been labeled as ‘read by $v_i$’ in line 12. Thus, next time Algorithm \ref{alg:ParallelDevice} gets to check if $q$ is read by $v_i$,  the condition in Line 11 evaluates to false.
\end{proof}

\begin{lemma}
\label{evenRead}
For any vertex $v_j$ that has an outgoing arc $(v_j, v_i)$, each path $q$ in $v_j.list$ will eventually be labeled `read by $v_i$' and will not change anymore.
\end{lemma}

\begin{proof}
Initially, each vertex $v_j$ includes $\{v_j\}$ as the only path, and this path is not read yet. The thread of $v_i$ executes the for-loop in Line 9 and iterates through all incoming arcs of each vertex $v_i \in V$ and will eventually get to $(v_j, v_i)$. As a result, any path $q \in v_j.list$ will be labeled on Line 12 because initially all paths are not read. If new paths are imported in  $v_j.list$ from its predecessors in subsequent iterations of the algorithm, then such paths will be labeled as `read by $v_i$' and will keep their status of being ‘read’.\end{proof}

\begin{algorithm}[H]
\caption{ExtendPath(Path[] $p$, Vertex $v$)}
    \begin{algorithmic}[1]
            \State Path[] $NewPath$ = $p$;
            \State $NewPath[5 + |p|$ + 1] = $v$;
            \State $NewPath[5] = p[5] + 1$; 
            \If {$v$ is the first vertex of $p$}
				\State $NewPath[4]$ = 1;     // CyclicPath flag
			\EndIf
            \State $NewPath[0]$ = 1;     // PathValidity flag  
			\State append $NewPath$ to $v.list$;
    \end{algorithmic}
\label{Function:ExtendPath}
\end{algorithm}

\begin{lemma}
\label{LocalflagFalse}
For each vertex $v_i \in V$, at some finite point in time, $v_i.LocalFlag$ will become false and will remain false.
\end{lemma}

\begin{proof}
Lemmas \ref{readOnce} and \ref{evenRead}  imply that all paths in the predecessors of $v_i$ will eventually be labeled as `read by the thread of $v_i$', and will keep their status of being `read'.  As such, the condition  on Line 11 will never become true again when processing arc $(v_j,v_i)$. Therefore, Algorithm \ref{alg:ParallelDevice} will no longer get to Line 23; i.e., $v_i.LocalFlag$ will become false (on Line 20) and will never become true again.
\end{proof}

\begin{lemma}
\label{[PublicflagFalse}
For any vertex $v_j$ that has an outgoing arc $(v_j, v_i)$, at some finite point in time, $v_j.PublicFlag$ will become false and will remain false.
\end{lemma}

\begin{proof}
If the graph has a vertex $v_k \in V$ with no incoming arcs, the loop in line 9 is skipped for $v_k$ by the thread of $v_k$ and $v_k.PublicFlag$ is set to false. Moreover, the thread of $v_k$ jumps over the loop in Line 28 for $v_k$ because $v_k.ReachedFrom$ is empty. Then, the PublicFlags of all immediate successors of $v_k$ are set to false by their corresponding threads. A similar phenomenon will propagate to all vertices reachable from $v_k$.
For a specific vertex $v_k \in V$ in a cycle, initially it reads its predecessor. This process of reading continues until each vertex has a list of paths with all vertices in the cycle (Lines 9 to 19). From this point on, no extension will occur for $v_k$ and $v_k.PublicFlag$ will remain false (Line 27).
\end{proof}

\begin{theorem}
\label{termination}
Algorithm \ref {alg:ParallelDevice} will eventually terminate.
\end{theorem}

\begin{proof}
Lemmas \ref{LocalflagFalse} and \ref{[PublicflagFalse} imply that at some finite point in time, $v_i.LocalFlag$ and $v_i.PublicFlag$ will become false for each $v_i \in V$ and will remain false. As such, when PublicFlag of all vertices in $v_i.ReachedFrom$ are set to false, the thread assigned to $v_i$ will eventually stop.When no more extensions occur for any vertex, the parallel algorithm terminates. 

\end{proof}

\subsection{Data Race Freedom }
In this section, we show that two neighboring threads/vertices do not cause any data races through accessing each other's path lists.  This is achieved without using any `atomic' instructions.

\begin{lemma}
\label{WriteHappensBF}
Let $(v_j, v_i)$ and $(v_i, v_k)$ be arcs in $A$. Adding a path from $v_j.list$ to the list of the vertex $v_i$ happens before reading such path from $v_i.list$ by the thread of vertex $v_k$. 

\end{lemma}

\begin{proof}
Thread of $v_k$ can read only the valid paths in vertex $v_i$ (line 11). The paths that are not yet fully written in the vertex $v_i$ (Line 14) are not readable.
\end{proof}

\begin{lemma}
\label{ReadHappensBF}
Let $(v_j, v_i)$ be an arc in $A$. Thread of $v_i$ reads each path  from $v_j.list$ in Line 11 before thread of $v_j$ removes such path from $v_j.list$ under conditions in Lines 3 and 4.
\end{lemma}

\begin{proof}
The thread of vertex $v_j$ can only remove a path $p$ from its list when p has been labeled as 'extended' and the corresponding successor flags have been set to true (Lines 3 and 4).

\end{proof}

\begin{theorem}
\label{data race free}
Algorithm \ref {alg:ParallelDevice} is data race free.
\end{theorem}

\begin{proof}
A program is data race free if all executions of the program are data race free. An execution is data race free if every pair of conflicting actions are related by the happens-before order. Lemmas \ref{WriteHappensBF} and \ref{ReadHappensBF} imply that Algorithm \ref {alg:ParallelDevice} guarantees happens-before order of every pair of conflicting actions by adjacent vertices on lines 3, 4, and 11.  

\end{proof}

\subsection{Completeness}

In order to show that TPGen generates all PPs, we present the following theorem.

\begin{theorem}
\label{data race free}
Algorithm \ref {alg:ParallelDevice} covers all PPs $100\%$ without a miss.
\end{theorem}
 
 \begin{proof}
 Complete (i.e.100\%) PPC can be proved through contradiction. Suppose that a maximal simple (prime) path $p(v_{i1},v_{in})$ is missed in $v_{in}.list$. Thus, the path $p(v_{i1},v_{in-1})$ will not be available in $v_{in-1}.list$. If we follow this rule in a backward direction toward $v_{i1}$, we reach the point that there is no vertex $v_{i1}$ in the list associated with this vertex. This is impossible because initially each vertex is on its list.
 \end{proof}

\section{Experimental Results}
\label{sec:ExperimantalResults}
This section presents the results of our experimental evaluations of the proposed GPU-based  method for PPs and TPs generation compared to the CPU-based approach proposed in \cite{fazli2019time}. The experimental benchmark consists of the set of ten CFGs from \cite{bang2015automatically} (which are taken from Apache Commons libraries). Table \ref{TableBenchMarks} presents the structure of these CFGs. Columns 2 to 5 of Table \ref{TableBenchMarks} provide the number of nodes, edges, and SCCs of each CFG. The total numbers of nodes and edges of all SCCs are mentioned as SccNodes and SccEdges, respectively. Columns 7 and 8 show the Cyclomatic Complexity (CC) and Npath complexities of the input CFGs. CC indicates the number of linearly independent complete paths of a given CFG \cite{mccabe1976complexity}. Npath complexity shows the number of execution paths while limiting the loops to at most one iteration of a CFG \cite{nejmeh1988npath}. To increase their structural complexity, we synthetically include extra nested loops and variety of conditional statements to create more SCCs. We compare the parallel and the sequential approaches with respect to their running time and memory consumption. The number of generated PPs for each CFG is provided in Column 9 of Table \ref{TableBenchMarks}.   We ran all the experiments on an Intel Core i7 machine with 3.6GHz X 8 processors and 16 GB of memory running Ubuntu 17.01 with gcc version 5.4.1. The parallel approach is implemented on a Nvidia GTX graphical processing unit equipped with 4G RAM and 768 CUDA cores.

\begin{table}
%\scriptsize
\caption{Structure of benchmark  input CFGs.}	
\begin{center}

\setlength\tabcolsep{3pt} % default value: 6pt
\begin{tabular}{|c|c|c|c|c|c|c|c|c|}
\hline
\multirow{2}{*}{\rotatebox[origin=c]{270}{  CFG}} & \multicolumn{6}{c|}{Graph Structure} & &\\ \cline{2-7} 
                              & {\rotatebox[origin=c]{270}{Nodes}} & {\rotatebox[origin=c]{270}{Edges}} & {\rotatebox[origin=c]{270}{SCC}} & {\rotatebox[origin=c]{270}{ SccNodes }} & {\rotatebox[origin=c]{270}{ SccEdges }} & CC  & \multirow{2}{*}{  Npath} & \multirow{2}{*}{ Prime Paths }     \\ \hline
1                             & 180   & 214   & 18   & 78       & 83       & 35  & $21244664*10^2$ & 35629                                  \\ \hline
2                             & 215   & 258   & 24   & 103      & 110      & 44  & $6115295232*10^2$ & 176481                                 \\ \hline
3                             & 244   & 290   & 27   & 115      & 125      & 49  & $339758624*10^4$ & 139684                                    \\ \hline
4                             & 355   & 431   & 38   & 160      & 173      & 68  &  $4508984864896*10^4$ & 253954                                 \\ \hline
5 & 486   & 567   & 47   & 223      & 244      & 104 & $115422132637413*10^{9}$ & 643738                                 \\ \hline
6                             & 723   & 853   & 75   & 331      & 351      & 131 &
$203282392447841*10^{18}$ & 1016762                           \\ \hline
7                             & 874   & 994   & 83   & 409      & 762      & 155 & 
$936725264399651*10^{25}$ & 1477397                 \\ \hline
8                             & 963   & 1119  & 93   & 448      & 490      & 213 & $13322314714616*10^{31}$ &  2573594                \\ \hline
9                             & 1441  & 1713  & 149  & 626      & 712      & 273 & $41323731079318*10^{55}$ & 4478382                     \\ \hline
10                            & 2160  & 2566  & 224  & 957      & 1073     & 404 & $840038691867497*10^{83}$ & 9563583                  \\ \hline
\end{tabular}
\end{center}
\label{TableBenchMarks}
\end{table}

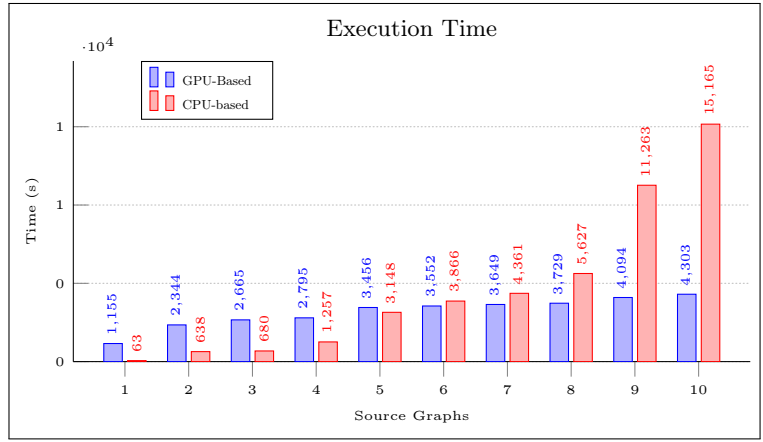
\begin{figure}[h]

\makebox[\textwidth][c]{
\fbox{\begin{minipage}[t]{0.8\textwidth}
\begin{tikzpicture}
\pgfplotsset{
    scale only axis,
   % scaled y ticks=base 5:0,
    ymin=0, ymax=16000,
    legend style={at={(.1,1)},anchor=north west,legend columns=1,column sep=2pt,nodes={scale=0.55, transform shape}},
    legend cell align={left},
	label style={font=\tiny},
    tick label style={font=\tiny},
    compat=1.16,
    ylabel style={font=\tiny},
    xlabel style={font=\tiny},
}
   \definecolor{myblue}{HTML}{1F7ED2}
  \begin{axis}[
    title  = Execution Time ,
    ylabel={Time (s)},
    xlabel={Source Graphs},
    ybar,
    bar width = 7pt,
    yticklabel={\pgfmathtruncatemacro\tick{\tick}\tick},
    x axis line style = { opacity = 1 },
	axis y line*=none,
    axis x line*=none,
    height			  = 4cm,
    width		      = 9cm,
    ymajorgrids,
    y grid style={densely dotted, line cap=round},
    enlarge x limits  = .09,
    enlarge y limits={upper,value=0.2},
    symbolic x coords = {1, 2, 3, 4, 5, 6, 7, 8, 9, 10},
    major tick length=.2cm,
    nodes near coords,
    xtick=data,
	every node near coord/.append style={rotate=90, anchor=west, font=\tiny},
  ]
  \addplot coordinates {(1,1155)(2,2344)(3,2665)(4,2795)(5,3456)(6,3552)(7,3649)(8,3729)(9,4094)(10,4303)};
  \label{plot_one}

  \addplot coordinates {(1,63)(2,638)(3,680)(4,1257)(5,3148)(6,3866)(7,4361)(8,5627)(9,11263)(10,15165)};  
  \label{plot_two}
   
  \legend{GPU-Based\hphantom{A},CPU-based}
  \end{axis}
\end{tikzpicture}
\end{minipage}}}
\caption{\small Time efficiency of CPU-based vs.  GPU-based method.}
\label{FigTimeComparison}
\end{figure}

Table \ref{TableComparision} presents the average execution time and maximum memory consumption by sequential and parallel approaches for each CFG. The reported timings for each approach is the average of twenty runs.   The bar graph of Figure \ref{FigTimeComparison} illustrates the time efficiency of the CPU-based and GPU-based approach. These values reflect the fact that the serial method running on the CPU has less consumption time for smaller CFGs than the parallel method implemented on the GPU. For the CFGs of the top five rows of Table \ref{TableBenchMarks}, on average, the CPU-based method consumed $61\%$ less time than the GPU-based method. On the other hand, for large CFGs at the bottom of Table \ref{TableBenchMarks}, the parallel method costs $39\%$ less time than the sequential method. This processing time efficiency increases significantly with increasing graph size. For example, the GPU-based time efficiency in the last graph is $71\%$. The recorded times indicate that by increasing the structural complexity, the GPU-based algorithm provides a better performance than the CPU-based algorithm. Therefore, for real-world applications that have a large number of lines and complex structures, the GPU-based algorithm is expected to be highly efficient.

\begin{table}
%\scriptsize
\caption{Time and memory costs of the GPU-based and CPU-based algorithms.}	
\begin{center}

\setlength\tabcolsep{3pt} % default value: 6pt
\begin{tabular}{|c|c|c|c|c|}
\hline
\multirow{2}{*}{CFG} & \multicolumn{2}{c|}{CPU-based}             & \multicolumn{2}{c|}{GPU-based} \\ \cline{2-5} 
                              & Memory (MB) & Time (s) & Memory (MB) & Time (s)   \\ \hline
1                             & 16.6937 & 63.69                      & 6.6885                & 1155.93               \\ \hline
2 &                            46.8456 & 638.06                   & 18.2185               & 2344.96               \\ \hline
3       &                     57.3789 & 680.98                      & 21.7319              & 2665.78               \\ \hline
4             &                 123.4179 & 1257.44                    & 44.1889  & 2795.83               \\ \hline
5 &  231.1347 & 3148.63                 & 82.6592              & 3456.33                \\ \hline
6 & 511.1835 & 3866.07                  & 227.2138             & 3552.35               \\ \hline
7 & 746.8261 & 4361.88                  & 318.8994            & 3649.31                \\ \hline
8    &                        906.5742 &  5627.34                 &     395.1738        & 3729.70                \\ \hline 
9 & 2029.2207 & 11263.36                 & 754.6743             & 4094.37               \\ \hline 10 & 4558.3593 & 15165.05                 & 1599.5354             & 4303.06              \\ \hline
\end{tabular}
\end{center}
\label{TableComparision}
\end{table}

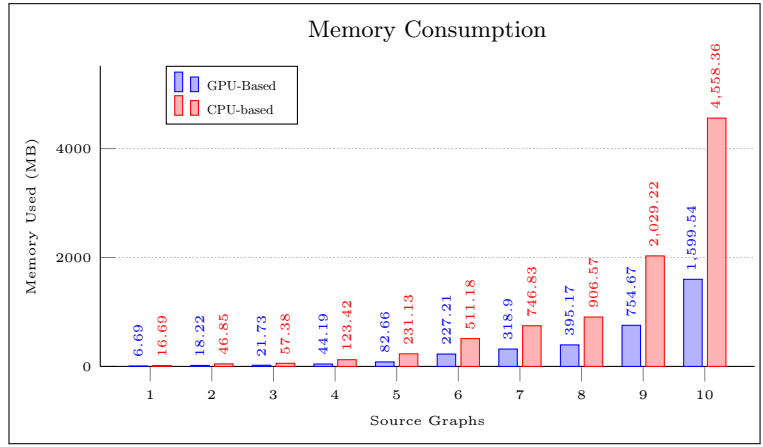
\begin{figure}[h]

\makebox[\textwidth][c]{
\fbox{\begin{minipage}[t]{0.8\textwidth}
\begin{tikzpicture}
\pgfplotsset{
    scale only axis,
   % scaled y ticks=base 5:0,
    ymin=0, ymax=4600,
    legend style={at={(.1,1)},anchor=north west,legend columns=1,column sep=2pt,nodes={scale=0.55, transform shape}},
    legend cell align={left},
	label style={font=\tiny},
    tick label style={font=\tiny},
    compat=1.16,
    ylabel style={font=\tiny},
    xlabel style={font=\tiny},
}
   \definecolor{myblue}{HTML}{1F7ED2}
  \begin{axis}[
    title  = Memory Consumption ,
    ylabel={Memory Used (MB)},
    xlabel={Source Graphs},
    ybar,
    bar width = 7pt,
    yticklabel={\pgfmathtruncatemacro\tick{\tick}\tick},
    x axis line style = { opacity = 1 },
	axis y line*=none,
    axis x line*=none,
    height			  = 4cm,
    width		      = 8.7cm,
    ymajorgrids,
    y grid style={densely dotted, line cap=round},
    enlarge x limits  = .09,
    enlarge y limits={upper,value=0.2},
    symbolic x coords = {1, 2, 3, 4, 5, 6, 7, 8, 9, 10},
    major tick length=.2cm,
    nodes near coords,
    xtick=data,
	every node near coord/.append style={rotate=90, anchor=west, font=\tiny},
  ]

  \addplot coordinates {(1,6.6885)(2,18.2185)(3,21.7319)(4,44.1889)(5,82.6592)(6,227.2138)(7,318.8994)(8,395.1738)(9,754.6743)(10,1599.5354)};  
  \label{plot_two}
  
  \addplot coordinates {(1,16.6937)(2,46.8456)(3,57.3789)(4,123.4179)(5,231.1347)(6,511.1835)(7,746.8261)(8,906.5742)(9,2029.2207)(10,4558.3593)};
  \label{plot_one}
   
  \legend{GPU-Based\hphantom{A},CPU-based}
  \end{axis}
\end{tikzpicture}
\end{minipage}}}
\caption{\small Space efficiency of CPU-based vs.  GPU-based method.}
\label{FigMemComparison}
\end{figure}

The bar graph of Fig.\ref{FigMemComparison} shows the space efficiency of the CPU-based and GPU-based approach. These values indicate that the GPU-based approach applying TPAM method has less memory requirements than the CPU-based method. On average, the GPU-based approach consumes $62\%$ less memory for the input CFGs. On the other hand, with the increase in the dimensions of the input CFGs, due to the contiguous allocation of memory, the CPU-based method has a lot of challenges in processing of very large real world CFGs.

\section{Related Work}
\label{sec:RelWorks}
This section discusses related works on the PPC in model based software testing context. Section \ref{sub:PathCov} presents the background of the path as well as PPC, and Section \ref{sub:GenPP} provides a high-level description of the PP generation algorithms presented in \cite{ammann2016introduction} \cite{fazli2019time}. Finally, Section \ref{sub:PathCov} discusses some related works to generate TPs and test data  for covering PPs. 
\subsection{Path Coverage}
\label{sub:PathCov}

The problem of finding minimum path for covering the sub paths of a digraph is a fundamental problem in some domains. Ntafos and Hakimi show that problems arising from practical program testing reduce to the minimum path cover problem \cite{ntafos1979path}. Given a directed graph $G(V,E)$, a path cover for the vertices of $G$ is a set of directed paths, $P = \{ p_1,p_2,...,  p_k\}$, such that for each $v_i \in V$, there is a $p_j \in P$ with $v_i \in p_j$ \cite{ahodata}. The minimum path cover problem consists of finding a path cover for G with the least number of paths. A minimum path cover consists of one path if and only if there is a Hamiltonian path in $G$. The Hamiltonian path problem is NP-complete, and hence the minimum path cover problem is NP-hard \cite{ahodata} \cite{aho1987efficient} .

In structural testing of a program, one must test all execution paths in the program. Even for small programs (with loops), the number of execution paths can be extremely large. This makes CPC testing an impossible task, since in the syntactic structure of the program's graph, there is an infinite number of paths to cover. A compromise is to test a subset of all executable paths of the PUT; such that its size is manageable and a significant amount of information, concerning program reliability, is obtained. This subset should cover interesting or potentially troublesome interactions among the code segments. The next best alternative to CPC, that can cover loops, is the PPC \cite{ammann2016introduction}. 

Ammann and Offutt defined the PPs test criterion to ensure strong coverage of loops without requiring an infinite number of execution paths \cite{ammann2016introduction}. The idea is to use simple paths only. However, even small programs can have a fairly large number of simple paths and we need to minimize the number of paths to consider. A simple path is a path without internal loops, but the path itself may be a loop. A PP is a maximal simple path that does not appear as a proper subpath of any other simple path. The excessive cost of enumerating all simple paths for a given program is avoided through only focusing on the maximal length simple paths. Since any path can be created by composing a group of simple paths, taking only the maximal length simple paths will ensure maximum simple path coverage. PPC subsumes a number of different coverage criteria that include data flow coverage and structural coverage. PPC is among the strongest coverage criteria that can be practically applied to software testing \cite{ammann2016introduction}.  This paper proposes a new parallel compositional method to generate PPs/TPs of an input CFG.

%firstly introduces PPC as a new program complexity metric. And then, % 

\subsection{Generating PPs}
\label{sub:GenPP}
Although generating PPs for large real-world programs is challenging, little effort has been made to introduce applicable approaches to generate PPs. To the best of our knowledge, there are two methods to generate PPs based on the syntactic structure of the PUT.  The first one is an edge-based algorithm proposed by Amman and Offutt \cite{ammann2016introduction} \cite{ammann2008graph}. The algorithm in the beginning considers to paths of length zero (nodes) as PP candidates. If each of them can extend by an edge holding PP condition, it extend to a path with length two, and so on. It will progressively continue the process until no path in the list can be extended. At the end of the algorithm, a pruning process will run. If any existing PP is covered by another one, it will be removed from the final PPs list.  The second one is a vertex based compositional method to generate the PPs proposed in \cite{fazli2019time}. The proposed compositional method (i) computes the component graph of the input control flow graph; (ii) calculates the set of PPs of each SCC, and (iii) generates the set of PPs of the input graph with very low time and space costs. The proposed method significantly outperforms the state-of-the-art in dealing with programs that have extremely large Npath complexities \cite{fazli2019time}. 

The proposed work in this paper is a parallel self-stabilizing vertex-based platform to expedite the PPs of the input CFG. It is an extension of the compositional vertex-based method in \cite{fazli2019time} to execute on a parallel processing platform based on the GPUs.

\subsection{Generating TPs}
\label{sub:GenTP}
Generating the set of TPs that cover all PPs of a given program requires each PP to be a subpath of at least one complete TPs. Once the set of all PPs is generated, the corresponding TPs touring them must be generated. Existing algorithms for discovering minimum TPs to cover a given set of paths fall in two categories, static and dynamic. 

Static methods are mainly based on the analysis of Control Flow Graph (CFG) of the PUT. Finding minimum path cover for the vertices (or edges) of a digraph has been studied earlier. Ntafos and Hakimi \cite{ntafos1979path} present two methods using SCC of the program's CFG based on Dilworth's theorem \cite{dilworth1950decomposition}. In one method the minimum path cover problem is transformed into a minimum flow problem; in the other it is transformed into a maximum matching problem. Aho and Lee \cite{aho1987efficient} tailor a minimum flow algorithm specifically for minimum TPss for node coverage. Li \textit{et~al.} \cite{li2012better} formulate the problem as a shortest superstring problem, which is NP-complete and then present two new solutions named Set-Covering Based and Matching-Based Prefix Graph to the problem. The proposed solutions use known approximation algorithms to solve the problem. Dwarakanath and Jankiti \cite{dwarakanath2014minimum} present an algorithm for the minimum flow problem based on the concept of Decreasing Paths. The concept is similar to that of the Augmenting Path in the Ford-Fulkerson algorithm \cite{ford2015flows} to find the maximum flow. Amman and Offutt present a simple solution for the generation of TPsTP that cover all PPs \cite{ammann2016introduction} . The solution starts with the longest PP and extends every PP to visit initial and final nodes thus forming a TPsTP. The process continues with the remaining uncovered longest PP. This algorithm does not attempt to minimize the number of TPs but is extremely simple.  Fazli and Afsharchi proposed a method based on merging, where obtain a set of complete TPs using internal PPs of all SCCs \cite{fazli2019time}. First, the set of SCC's entry exit paths that cover all internal PPs of all SCCs is generated. Then, the method in \cite{fazli2019time} merges these paths with main component graph complete paths, thereby yielding complete TPs that cover all incomplete PPs. 

Dynamic methods are based on execution of an instrumented version of the PUT and analysis of covered level against a set of desired paths. To generate test cases for covering a specific set of paths, a number of algorithms have been proposed. Hosseini and Jalili \cite{hoseini2014automatic} present a genetic algorithm to generate TPs covering all PPs. Other nature inspired algorithms have also been proposed, for example, Sayyari and Emadi \cite{sayyari2015automated} propose a solution based on ant colony optimization algorithm to develop TPs.  Srivastava \textit{et~al.} \cite{srivastava2010optimized} propose a technique using ant colony optimization to generate optimized test sequences from markov chain model of the software under test. Bidgoli \textit{et~al.} \cite{bidgoli2017using} utilize two prominent swarm intelligence algorithms, i.e., ACO and PSO, along with a normalized fitness function to provide an approach for covering PPs. Lin \textit{et~al.} \cite{lin2001automatic} and Bueno\textit{et~al.} \cite{bueno2002automatic} introduce methods for test data generation based on the GA algorithm with covering PPs capability. We summarize the reviewed related works relevant to PPs/TPs generation literature in Table \ref{table:RelatedWorks}.   

The parallel processing platform Proposed in this paper is abased on the GPUs to extract the PPs/TPs. It belongs to static methods category as it is based on the CFG of the PUT. Most of real world programs even with just a few statements, have a large number of PPs. Processing time grows exponentially as the number of conditional and loop statements grows. This limitation is a big challenge for the efficiency and scalability of the existing PP generation algorithms for large real world programs. However, our technique overcomes this limitation by modularized analysis of the CFG. It generates PPs of SCCs as well as the component graph's PPs separately and produces the rest of the PPs by integrating the existing generated PPs. This method can be highly scalable by significantly reducing processing time requirements.

%\begin{landscape}

\begin{table}[h]
\scriptsize
    \centering
    \caption{Related works on PPs and TPs generation}
    \begin{tabular}{|p{1.45 cm}|l|l|l|}
    \hline
        Method & Work & Criterion & Procedure \\ \hline
        \multirow{2}{*} {\begin{tabular}{@{}c@{}}Static Path \\ Generator\end{tabular}} & Amman et al \cite{ammann2016introduction} & PPs Generation & Edge based \\ \cline{2-4} 
        ~ & Fazli et al \cite{fazli2019time} & PPs Generation & Compositional vertex based \\ \hline
       \multirow{6}{*} {\begin{tabular}{@{}c@{}}Static  \\ Test Path \\Generator\end{tabular}} & Amman et al \cite{ammann2016introduction} & PP Coverage & Extend PP to visit first/end vertices \\ \cline{2-4} 
        ~ & Fazli et al \cite{fazli2019time} & PP Coverage & Transform/Merge internal PP of all SCCs \\ \cline{2-4} 
        ~ & Ntafos et al \cite{ntafos1979path} & Edge Coverage & Minimum flow and maximum matching \\ \cline{2-4} 
        ~ & Aho et al \cite{aho1987efficient} & Node Coverage & Minimum flow \\ \cline{2-4} 
        ~ & Li et al \cite{li2012better} & Edge Coverage & Set-Covering/Matching-Based Prefix Graph \\ \cline{2-4} 
        ~ & Dwarakanath et al \cite{dwarakanath2014minimum}& PP Coverage & Minimum flow based on Decreasing Paths \\ \hline
       \multirow{6}{*} {\begin{tabular}{@{}c@{}}Dynamic  \\ Test Path \\ Covering \end{tabular}} & Hosseini et al \cite{hoseini2014automatic} & PP Coverage & Genetic algorithm \\ \cline{2-4} 
        ~ & Sayyari et al \cite{sayyari2015automated} & PP Coverage & Ant colony optimization \\ \cline{2-4} 
        ~ & Srivastava et al \cite{srivastava2010optimized} & PP Coverage & Ant colony optimization \\ \cline{2-4} 
        ~ & Bidgoli et al \cite{bidgoli2017using} & PP Coverage & Ant colony and PSO \\ \cline{2-4} 
        ~ & Lin et al \cite{lin2001automatic} & PP Coverage & Genetic algorithm \\ \cline{2-4} 
        ~ & Bueno et al \cite{bueno2002automatic} & PP Coverage & Genetic algorithm \\ \hline
    \end{tabular}
    \label{table:RelatedWorks}
\end{table}
%\end{landscape}

\section{Conclusions and Future Work}
\label{sec:conclude}

We  presented a novel scalable GPU-based method,  called TPGen,   for the generation of all Prime Paths (PPs) and Test Paths (TPs) used in structural testing and in test data generation.  TPGen outperforms existing methods for PPs and TPs generation in several orders of magnitude, both in time and space efficiency. To reduce TPGen's memory costs, we designed a non-contiguous and hierarchical memory allocation method, called Three-level Path Access Method (TPAM), that enables efficient storage of maximal simple paths in memory. TPGen does not use any synchronization primitives for the execution of the kernel threads on GPU, and starting from any execution order of threads, TPGen generates the PPs ending in any individual vertex; hence providing a fully asynchronous self-stabilizing GPU-based algorithm.  As an extension of this work, we will expand the proposed benchmark with more structurally complex programs. We also will integrate PPs/TPs generation with constraint solvers towards generating test data in programs with high structural complexity.

%In addition to TPGen, we developed (i) a new measure for the structural complexity of programs based on the notion of prime paths, and (ii) a set of benchmark programs for structural complexity. 

 \bibliographystyle{splncs04}
  \bibliography{biblio.bib}

\begin{thebibliography}{10}
\providecommand{\url}[1]{\texttt{#1}}
\providecommand{\urlprefix}{URL }
\providecommand{\doi}[1]{https://doi.org/#1}

\bibitem{ahodata}
Aho, A.V., Hopcroft, J.E., Ullman, J.D.: Data structures and algorithms (1983)

\bibitem{aho1987efficient}
Aho, A.V., Lee, D.: Efficient algorithms for constructing testing sets,
  covering paths, and minimum flows. AT\&T Bell Laboratories Tech. Memo.
  CSTR159 pp. 1--15 (1987)

\bibitem{ammann2016introduction}
Ammann, P., Offutt, J.: Introduction to software testing. Cambridge University
  Press (2016)

\bibitem{ammann2008graph}
Ammann, P., Offutt, J., Xu, W., Li, N.: Graph coverage web applications (2008)

\bibitem{bang2015automatically}
Bang, L., Aydin, A., Bultan, T.: Automatically computing path complexity of
  programs. In: Proceedings of the 2015 10th Joint Meeting on Foundations of
  Software Engineering. pp. 61--72. ACM (2015)

\bibitem{bidgoli2017using}
Bidgoli, A.M., Haghighi, H., Nasab, T.Z., Sabouri, H.: Using swarm intelligence
  to generate test data for covering prime paths. In: International Conference
  on Fundamentals of Software Engineering. pp. 132--147. Springer (2017)

\bibitem{bueno2002automatic}
Bueno, P.M.S., Jino, M.: Automatic test data generation for program paths using
  genetic algorithms. International Journal of Software Engineering and
  Knowledge Engineering  \textbf{12}(06),  691--709 (2002)

\bibitem{dilworth1950decomposition}
Dilworth, R.P.: A decomposition theorem for partially ordered sets. Annals of
  Mathematics pp. 161--166 (1950)

\bibitem{dwarakanath2014minimum}
Dwarakanath, A., Jankiti, A.: Minimum number of test paths for prime path and
  other structural coverage criteria. In: IFIP International Conference on
  Testing Software and Systems. pp. 63--79. Springer (2014)

\bibitem{eisenstat1981algorithms}
Eisenstat, S.C., Schultz, M.H., Sherman, A.H.: Algorithms and data structures
  for sparse symmetric gaussian elimination. SIAM Journal on Scientific and
  Statistical Computing  \textbf{2}(2),  225--237 (1981)

\bibitem{fazli2019time}
Fazli, E., Afsharchi, M.: A time and space-efficient compositional method for
  prime and test paths generation. IEEE Access  \textbf{7},  134399--134410
  (2019)

\bibitem{ford2015flows}
Ford~Jr, L.R., Fulkerson, D.R.: Flows in networks. Princeton university press
  (2015)

\bibitem{hoseini2014automatic}
Hoseini, B., Jalili, S.: Automatic test path generation from sequence diagram
  using genetic algorithm. In: Telecommunications (IST), 2014 7th International
  Symposium on. pp. 106--111. IEEE (2014)

\bibitem{li2012better}
Li, N., Li, F., Offutt, J.: Better algorithms to minimize the cost of test
  paths. In: Software Testing, Verification and Validation (ICST), 2012 IEEE
  Fifth International Conference on. pp. 280--289. IEEE (2012)

\bibitem{lin2001automatic}
Lin, J.C., Yeh, P.L.: Automatic test data generation for path testing using
  gas. Information Sciences  \textbf{131}(1-4),  47--64 (2001)

\bibitem{mccabe1976complexity}
McCabe, T.J.: A complexity measure. IEEE Transactions on software Engineering
  (4),  308--320 (1976)

\bibitem{nejmeh1988npath}
Nejmeh, B.A.: Npath: a measure of execution path complexity and its
  applications. Communications of the ACM  \textbf{31}(2),  188--200 (1988)

\bibitem{ntafos1979path}
Ntafos, S.C., Hakimi, S.L.: On path cover problems in digraphs and applications
  to program testing. IEEE Transactions on Software Engineering (5),  520--529
  (1979)

\bibitem{sayyari2015automated}
Sayyari, F., Emadi, S.: Automated generation of software testing path based on
  ant colony. In: Technology, Communication and Knowledge (ICTCK), 2015
  International Congress on. pp. 435--440. IEEE (2015)

\bibitem{srivastava2010optimized}
Srivastava, P.R., Jose, N., Barade, S., Ghosh, D.: Optimized test sequence
  generation from usage models using ant colony optimization. International
  Journal of Software Engineering \&Applications  \textbf{2}(2),  14--28 (2010)

\bibitem{tarjan1972depth}
Tarjan, R.: Depth-first search and linear graph algorithms. SIAM journal on
  computing  \textbf{1}(2),  146--160 (1972)

\end{thebibliography}

\end{document}